\documentclass[letterpaper,twocolumn,10pt]{article}

\usepackage{template/usenix-2020-09}

\usepackage[normalem]{ulem}
\usepackage{tikz}
\usepackage{comment}
\usepackage{algorithm}
\usepackage{algpseudocode}
\usepackage{graphicx}
\usepackage{textcomp}
\usepackage{enumitem}
\usepackage{xspace}
\usepackage{multirow}
\usepackage{comment}
\usepackage{minted}
\usepackage{tcolorbox}
\usepackage{xcolor}
\usepackage{multicol}
\usepackage{threeparttable}
\usepackage{xstring}
\usepackage{booktabs} 
\usepackage{filecontents}
\usepackage{wrapfig}
\usepackage{amsmath}
\usepackage{cancel}
\usepackage{soul}

\begin{document}

\date{}

\author{
{\rm Xi Wang}\\
University of California, Merced
\and
{\rm Jie Liu}\\
University of California, Merced
\and
{\rm Shuangyan Yang}\\
University of California, Merced
\and
{\rm Jongryool Kim}\\
SK hynix
\and
{\rm Pengfei Su}\\
University of California, Merced
\and
{\rm Dong Li}\\
University of California, Merced
} 

\definecolor{dong}{RGB}{0,0,200}
\definecolor{check}{RGB}{0,0,0}
\definecolor{sherry}{RGB}{255,140,0}
\definecolor{draft}{RGB}{102,204,0}

\newcommand{\name}{PTMT\xspace}
\title{Hybrid Adaptive Tuning for Tiered Memory Systems}

\maketitle

\begin{abstract}
Memory tiering provides a cost-effective solution to increase memory capacity, utilization, and even bandwidth. Memory tiering relies on system software for memory profiling, detection of frequently accessed pages, and page migration. Such a system software often comes with system parameters. The configurations of those parameters impact application performance. We comprehensively classify system parameters, and characterize the sensitivity of application performance to them using representative memory tiering solutions. Furthermore, we introduce a lightweight and user-friendly framework \name, which automates tuning of parameters at runtime for various memory tiering solutions. We identify major challenges for online tuning of memory tiering. \name uses a hybrid ``offline + online'' tuning method: while the offline phase builds a performance database for online queries and reduces runtime overhead, the online phase uses reinforcement learning (customized to memory tiering) to tune. \name improves performance by \textcolor{check}{30\%, 26\%, 21\%, and 14\%}, on four memory tiering solutions (TPP, UPM, Colloid, and Auto\-NUMA), compared to using the default configurations. \name outperforms the state-of-the-art \cite{idt2024chang} by \textcolor{check}{32\%} on average.


\end{abstract}%

\section{Introduction}


Memory tiering provides a cost-effective solution to increase memory capacity, utilization, and even bandwidth. By integrating multiple memory components with various properties (e.g., capacity, latency, bandwidth, and monetary cost), memory tiering addresses the scaling challenges faced by the traditional single-tiered memory.


A typical tiered-memory system comprises fast memory and slow memory where fast memory offers shorter latency or higher bandwidth but limited memory capacity, and slow memory provides larger capacity but relatively worse performance. To manage tiered memory, a system software can enable a unified address space across memory components and model each memory component as a  NUMA node ~\cite{meta_tpp,Memtis2023,Hemem2021,eurosys24:mtm,atc24_hm}. Such a system software (called the memory tiering solution) 
relies on dynamic memory profiling to collect statistics on page access frequency and recency, and detect hot pages. Based on the profiling and detection results, the memory tiering solution allocates or migrates frequently accessed memory pages (hot pages) into fast memory for high performance (called page promotion), while placing less frequently accessed pages (cold pages) into slow memory to save fast memory space (called page demotion). 

\textbf{Problems.} 
The memory profiling, hotness detection, and page migration in memory tiering often come with system parameters. Those parameters control memory profiling overhead and quality, and impact page migration frequency and effectiveness. The parameter configuration impacts the performance significantly. For example, \textcolor{check}{Colloid~\cite{colloid2024} \hspace{3pt} (a state-of-the art \hspace{3pt} memory tiering \hspace{3pt} solution) \hspace{3pt}  uses \hspace{3pt}  a parameter,  \\ \texttt{watermark\_scale\_factor}, to control page demotion aggressiveness and maintain available fast-memory space for page promotion. Using an appropriate configuration for this parameter improves Graph500~\cite{graph500} performance by 17\%, compared to using the default configuration \textcolor{check}{(see Sec.  \ref{sec:sys_perf_senstivity})}.}


However, tuning system parameters for memory tiering is challenging, because different workloads, either coming from different execution phases of the same application or different applications, 
can exhibit different memory access patterns, hence demanding different parameter configurations. Existing memory tiering solutions \textcolor{check}{~\cite{Memtis2023,atc24_hm,eurosys24:mtm,Hemem2021,autonuma_tierd_memory,meta_tpp,tiering0.8, AutoTiering2021, Nomad2024, Google_TMTS2023, MULTI-CLOCK2022, HotBox2022bergman, colloid2024, song2025hybridtier, huaicheng2025tiered, cmpi2025, xu2026cccl, wang2025performance, ren2025machine}} use invariant configurations, 
which leads to suboptimal performance. 
Today, parameter tuning for memory tiering remains largely manual, as no systematic approach exists to automate this process across different tiering solutions.

We make three key \textbf{insights} driving our work. 
First, tuning system parameters for memory tiering is a inherently ``stateful'' problem. 
The effect of a system parameter setting depends not only on workload characteristics (e.g., cache hit rates and arithmetic intensity) but also on memory tiering states.
The memory tiering state refers to how pages have been allocated and migrated by a memory tiering solution. 
A workload under different memory tiering states can respond differently to the parameter tuning, even if the workload has the same characteristics. 

Second, the combination of many parameter configurations and the essentially unbounded space of workload states creates a rather large search space to decide the optimal configuration. 
Building a \textit{general} model aiming to navigate the space for all applications can be 
ineffective to decide the configurations.

Third, considering that characteristics and memory-tiering states of a workload evolve over time, the problem naturally lends itself to reinforcement learning (RL), which is designed to adapt to changing environments and learn effective decisions online.

In this paper, we introduce \underline{P}arameter \underline{T}uning for \underline{M}emory \underline{T}iering (\name), a framework that enables automatic tuning of system parameters across diverse memory tiering solutions. \name faces multiple challenges discussed as follows.


\textbf{Challenges.} 
First, the process of deciding parameter configurations must be lightweight to make runtime-tuning feasible. In memory tiering, the time intervals for periodical memory profiling and page migration are typically every few seconds~\cite{doudali2019kleio, doudali2022cronus, doudali2021cori, idt2024chang, AMP2020} to accommodate changing memory access patterns in applications. Consequently, the time overhead of deciding parameter configurations must be on the same scale or even smaller. Exhaustive search or frequent sampling of the search space is therefore not feasible.

Second, although RL is promising for adapting to changing environments and deciding parameter configurations, RL convergence can be slow due to the large state and action space inherent in memory tiering. How to accelerate RL convergence to timely respond to changing memory access patterns is challenging.



Third, parameter tuning must account for how different application inputs shape memory access behavior. Inputs often influence an application's control-flow paths, which in turn affect memory-level parallelism and ultimately the frequency and recency of memory accesses. Efficiently capturing these input-dependent effects at runtime is challenging.




\textbf{Solutions.} To address the above challenges, \name uses a hybrid ``offline + online'' tuning approach. The offline phase, in essence, builds a ``performance database'' where there are a number of workload states associated with parameter configurations and aftermath performance. During the online phase, the performance database is queried using the application's current workload state (WS) to find a parameter configuration that leads to the best performance on a similar WS. To handle a diverse set of application inputs that lead to WS outliers, \name uses an RL model that automatically learns a policy to decide the parameter configurations.

Using the hybrid approach, the time overhead of deciding the parameter configuration is small. The offline phase clusters WSs by similarity, such that finding a group of similar workloads at runtime is just a matter of finding a nearest centroid among those clusters. Within the cluster, WSs are further organized and indexed to enable efficient lookup of the most similar WSs. In addition, using the performance database, we build a pre-trained model as the RL model such that the model convergence can be faster. 

We adopt an application-specific solution to design \name, meaning both the performance database and the RL model are tailored to each application. 
This specialization enables \name to identify higher-quality parameter configurations, yielding a 20\% performance gain while also reducing model-construction cost by $11\times$ on average, compared to a general approach (see Sec. \ref{sec:eval_rl}). 

\textbf{Contributions.} We summarize the main contributions as follows.

\begin{itemize}[leftmargin=*,noitemsep,topsep=0pt]
    \item We comprehensively classify system parameters and characterize the sensitivity of application performance to them using four representative memory tiering solutions (Auto\-NUMA \cite{autonuma_tierd_memory}, Colloid ~\cite{colloid2024}, TPP~\cite{meta_tpp}, and UPM~\cite{PTMT_UPM}).

    \item We identify key technical challenges in enabling automatic tuning of system parameters for memory tiering. We introduce \name, which is the first lightweight and user-friendly parameter-tuning framework for memory tiering. 
    \textcolor{check}{\name requires no changes to applications or the operating system (OS) and can be applied across diverse memory-tiering solutions.}
    The design of \name answers multiple questions to tune memory tiering and addresses unique challenges unseen in any other tuning frameworks, e.g., how to avoid OS modification for memory tiering? Should we use an application-specific solution or a general solution for memory tiering?

    \item By adaptively tuning system parameters, \name achieves \textcolor{check}{14\%, 21\%, 30\%, and 26\%} performance improvements on AutoNUMA, \textcolor{check}{Colloid}, TPP, and UPM respectively, compared to using the default parameter configurations. Compared with the state-of-the-art parameter-tuning approach~\cite{idt2024chang}, \name delivers a \textcolor{check}{32\%} performance improvement on average. 
\end{itemize}

\section{Understanding Parameter Tuning in Memory Tiering}

\begin{table*}[!t]  
\caption{Memory tiering solutions and their system parameters.}  
\label{tab:page_management}  
\footnotesize  
\resizebox{\textwidth}{!}{  
\begin{tabular}{|c|c|c|c|c|c|c|}  
\hline  
\multirow{2}{*}{\textbf{Solutions}} &  
  \multicolumn{2}{c|}{\textbf{Memory Profiling}} &  
  \multicolumn{2}{c|}{\textbf{Hotness Detection}} &  
  \multicolumn{2}{c|}{\textbf{Page Migration}} \\ \cline{2-7}   
 &  
  \textbf{Mechanism} &  
  \textbf{Parameters} &  
  \textbf{Mechanism} &  
  \textbf{Parameters} &  
  \textbf{Mechanism} &  
  \textbf{Parameters} \\ 
\hline
\addlinespace[4pt] 
\hline
AutoNUMA &  
  \begin{tabular}[c]{@{}l@{}}NUMA \\Scanning\\ + Hint Fault\end{tabular} &  
  \begin{tabular}[c]{@{}l@{}}\texttt{scan\_size\_mb} \\ \texttt{scan\_period\_max\_ms} \\ \texttt{scan\_period\_min\_ms} \\ \texttt{scan\_delay\_ms}\end{tabular} &  
  Hint Fault Latency &  
  \begin{tabular}[c]{@{}l@{}}\texttt{hot\_threshold} \end{tabular} &  
  \begin{tabular}[c]{@{}c@{}}\textcolor{check}{NUMA Hint Fault}\\kswapd\end{tabular} &  
  \begin{tabular}[c]{@{}l@{}}\texttt{watermark\_scale\_factor} \\ \texttt{promote\_rate\_limit} \end{tabular}  \\ \hline  
\textcolor{check}{Colloid} &  
  \begin{tabular}[c]{@{}l@{}}NUMA \\Scanning\\ + Hint Fault\end{tabular} &  
  \begin{tabular}[c]{@{}l@{}}\texttt{scan\_size\_mb} \\ \texttt{scan\_period\_max\_ms} \\ \texttt{scan\_period\_min\_ms} \\ \texttt{scan\_delay\_ms}\end{tabular} &  
  \begin{tabular}[c]{@{}c@{}}Hint Fault Latency \\ \textcolor{check}{Fast/Slow Mem Latency Balancing}\end{tabular} &  
  \begin{tabular}[c]{@{}l@{}}\texttt{hot\_threshold} \end{tabular} &  
  \begin{tabular}[c]{@{}c@{}}\textcolor{check}{NUMA Hint Fault}\\kswapd\end{tabular} &  
  \begin{tabular}[c]{@{}l@{}}\texttt{watermark\_scale\_factor} \\ \texttt{promote\_rate\_limit} \end{tabular} \\ \hline  
TPP &  
  \begin{tabular}[c]{@{}l@{}}NUMA \\Scanning\\ + Hint Fault\end{tabular} &  
  \begin{tabular}[c]{@{}l@{}}\texttt{scan\_size\_mb} \\ \texttt{scan\_period\_max\_ms} \\ \texttt{scan\_period\_min\_ms} \\ \texttt{scan\_delay\_ms}\end{tabular} &  
  \begin{tabular}[c]{@{}c@{}}PTE Scanning \\ LRU-based Active List\end{tabular} &  
  \texttt{None} &  
  \begin{tabular}[c]{@{}c@{}}\textcolor{check}{NUMA Hint Fault}\\kswapd\end{tabular} &  
  \begin{tabular}[c]{@{}l@{}}\texttt{watermark\_scale\_factor} \\ \texttt{demote\_scale\_factor}\end{tabular} \\ \hline  
\multirow{2}{*}{UPM} &  
  \begin{tabular}[c]{@{}l@{}}HW Event \end{tabular} &  
  \multirow{2}{*}{\texttt{sample\_period}} &  
  \begin{tabular}[c]{@{}l@{}}Sampled Access \end{tabular} &  
  \multirow{2}{*}{\texttt{hot\_threshold}} &  
  \begin{tabular}[c]{@{}l@{}}move\_pages()\end{tabular} &  
  \multirow{2}{*}{\textcolor{check}{\texttt{page\_migration\_interval}}} \\   
  & Sampling & & Count & & syscall & \\ \hline  
\end{tabular}  
}
\end{table*}
\vspace{-10pt}

\subsection{Memory Profiling}
\label{sec:mem_prof}
There are three common memory profiling methods.

\textbf{PTE scanning-based profiling} tracks page accesses by manipulating \texttt{ACCESS} bits in page table entries (PTEs). By repeatedly scanning PTE to detect the change of \texttt{ACCESS} bit, we can detect memory accesses. This profiling method is used in Linux's memory swapping mechanism to maintain active and inactive lists for tracking page hotness. This method is commonly used in memory tiering~\cite{Google_TMTS2023, heterovisor2015, heteroos2017, daos2022, MULTI-CLOCK2022, eurosys24:mtm, DAMON}.

\textbf{NUMA hint faults-based profiling} 
employs NUMA scanning that partitions the whole virtual address space into memory regions, and profile them in order by manipulating \texttt{PRESENT} bit. With the bits set in a memory region, any page access in the region triggers a hint page-fault, which allows the kernel to identify access recency. After a time interval, NUMA scanning moves on to another memory region to profile. This profiling method is used in recent efforts on memory tiering~\cite{autonuma_tierd_memory, tiering0.8, meta_tpp, HotBox2022bergman, AutoTiering2021, Nomad2024, colloid2024}.

\textbf{Performance counter-based profiling} tracks memory accesses by sampling hardware performance counters available in modern CPUs, which allows the user to track memory accesses at various granularities (e.g., cache line level and page level). Compared to the other two profiling methods, this method can be more lightweight. It is commonly used in the existing efforts~\cite{TMP2021, Google_TMTS2023, Hemem2021, MaPHeA2021, Memtis2023}.

\subsection{Memory Tiering}
\label{sec:mt}

A memory tiering solution typically includes three components~\cite{Memtis2023,atc24_hm,eurosys24:mtm,Hemem2021,autonuma_tierd_memory,meta_tpp,tiering0.8, AutoTiering2021, Nomad2024, Google_TMTS2023, MULTI-CLOCK2022, HotBox2022bergman, colloid2024, song2025hybridtier, huaicheng2025tiered} (i.e., memory profiling, hotness detection, and page migration), each of which includes one or more system parameters. Table~\ref{tab:page_management} summarizes the parameters in multiple representative memory-tiering solutions. We discuss them as follows.



\textbf{AutoNUMA} \cite{autonuma_tierd_memory} is the default memory tiering solution in Linux. It utilizes NUMA hint faults to decide page promotion. AutoNUMA identifies page hotness based on the hint fault latency, which is the time interval between the recent hint fault time and the recent scanning time of a page. When the hint fault latency is short enough, the page is promoted. AutoNUMA relies on Linux swapping daemon (\texttt{kswapd}) for page demotion. When fast memory is in shortage, \texttt{kswapd} uses PTE scanning to determine cold pages to demote.

\textbf{Colloid} \cite{colloid2024} is a state-of-the-art memory tiering solution. Integrated with AutoNUMA (Linux v6.3), Colloid prevents overloading of fast memory. Like AutoNUMA, Colloid uses NUMA hint faults and fault latency to identify page hotness and promotion. Different from AutoNUMA, Colloid launches a background kernel thread to monitor memory access latencies for both fast and slow memories. Upon NUMA hint faults, Colloid determines the migration direction — whether to promote or demote — by comparing with expected access latencies.




\textbf{TPP} \cite{meta_tpp} relies on NUMA scanning to trigger page hint faults. However, unlike AutoNUMA and \textcolor{check}{Colloid}, TPP identifies hint-faulted pages as hot by verifying if these pages are present in the OS’s active list. Furthermore, TPP decouples allocation from demotion by performing demotion asynchronously in a background process. 


\textbf{UPM} \cite{PTMT_UPM} is a user-space page management solution developed by us. UPM uses performance counters to track memory accesses at the page level. 
While there are other performance counter-based solutions (e.g., HeMem~\cite{Hemem2021} and MEMTIS~\cite{Memtis2023}), we cannot use them due to stability issues 
in our evaluation. UPM shares a similar design with HeMem in memory profiling and hotness detection. 




\subsection{System Parameters and Performance Sensitivity}
\label{sec:sys_perf_senstivity}
We classify system parameters in the context of the three components in memory tiering. 

\textbf{System parameters in memory profiling} are often used to set a balance between profiling overhead and quality. Take NUMA scanning as an example. Frequent NUMA scanning over memory regions leads to large profiling overhead because of frequent hint faults and manipulation of PTE bits. However, frequent NUMA scanning can capture page access in a timely manner. Hence, Linux provides a system parameter, \texttt{scan\_size\_mb}, allowing users to strike the above balance. This parameter controls the size of the memory region to scan. Given a scanning time interval, 
a smaller region to scan leads to smaller overhead but suffers from lower profiling quality, and vice versa. 

\textbf{System parameters in hotness detection} are used to control the definition of page hotness. For example, UPM uses a system parameter \texttt{hot\_threshold}. The number of accesses to a page is compared against it to determine page hotness. A smaller \texttt{hot\_threshold} increases the number of candidate pages to promote but also raises migration overhead, and vice versa. UPM does not change \texttt{hot\-\_threshold} throughout application execution. AutoNUMA and \textcolor{check}{Colloid} have a similar system parameter, but dynamically adjust it according to a promotion rate limit. \name does not tune such a dynamic parameter, and focuses on static ones.

\textbf{System parameters in page migration} can be used to balance fast-memory utilization and migration overhead. For page promotion, AutoNUMA and \textcolor{check}{Colloid} use \texttt{promote\_} \texttt{rate\_limit} to control promotion traffic. For page demotion, AutoNUMA, \textcolor{check}{Colloid}, and TPP employ \texttt{kswapd} to demote pages in the inactive list. These three solutions \textit{periodically} inspect free space in fast memory and wake up \texttt{kswapd} when the free memory falls below a threshold \texttt{watermark\_scale\_} \texttt{factor}. TPP further introduces a threshold, \texttt{demote\_scale\_} \texttt{factor}, to determine when 
the background demotion process should terminate. 
\textcolor{check}{A higher \texttt{watermark\_\allowbreak scale\_\allowbreak factor} or \texttt{demote\_\allowbreak scale\_\allowbreak factor} enables more prompt page demotion, creating more space for page promotion but at the cost of higher page migration overhead, and vice versa. }

Unlike the other three memory-tiering solutions, UPM uses \texttt{page\_migration\_interval} to control page migration frequency. Specifically, UPM performs page demotion and promotion at each \texttt{page\_\allowbreak migration\_interval}. Instead of using \texttt{kswapd} for demotion, UPM uses its own thread to demote pages whose access count falls below \texttt{hot\_\allowbreak threshold}. This page demotion continues until there is enough fast memory available for hot page promotion. A smaller \texttt{page\_\allowbreak migration\_interval} improves responsiveness to the change of hot pages but incurs higher overhead, and vice versa.

\begin{figure}[t!]
	\centering
        \includegraphics[width=1.0\linewidth]{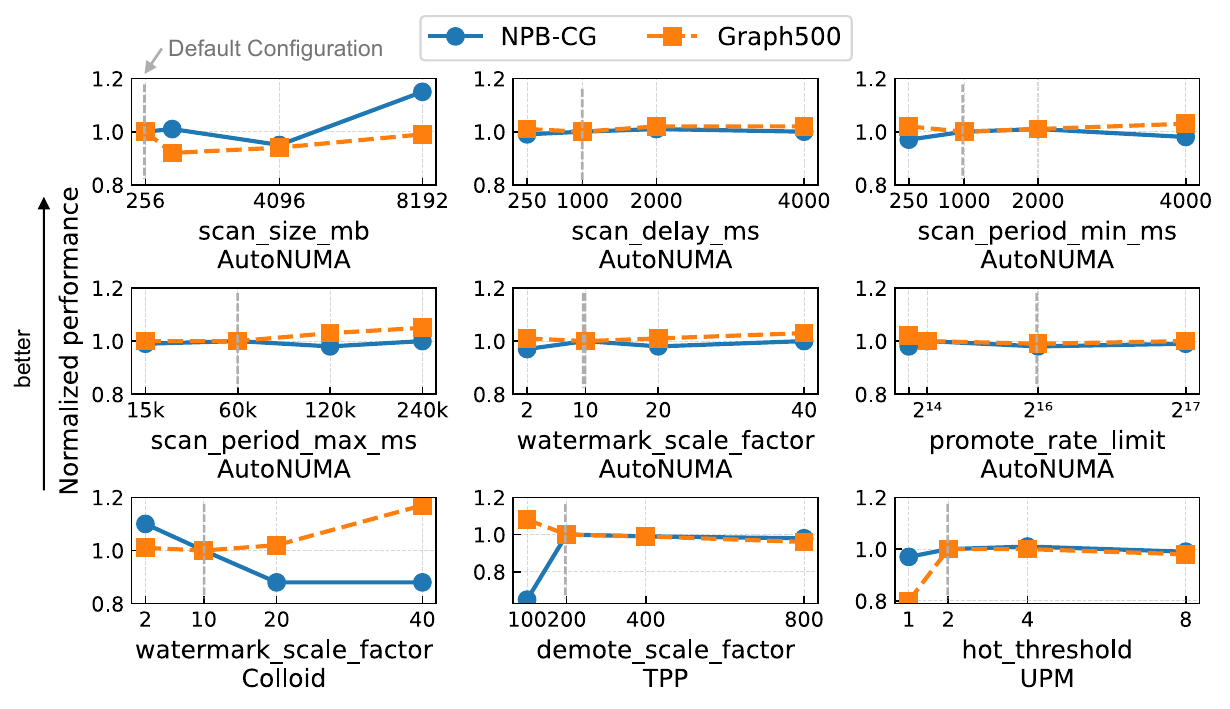}
    \vspace{-10pt}

	\caption{Performance with various parameter configurations, normalized by that with the default configurations.}
	\label{fig:parameter_sensitivity} 
\end{figure}

\textbf{Performance sensitivity study.} 
We study performance sensitivity to parameter configurations using representative memory tiering solutions discussed in Sec.~\ref{sec:mt}. We study the parameters listed in Table \ref{tab:page_management}, covering memory profiling, hotness detection, and page migration. 
We run two benchmarks, CG from the NAS Parallel Benchmarks (NPB) suite~\cite{NASA_NPB} and Graph500 ~\cite{graph500}, using 24 threads on  an Intel Optane machine (depicted in Sec.~\ref{sec:eval_method}).  We use static tuning, i.e., the parameters are invariant during each execution.  The results are presented in Figure~\ref{fig:parameter_sensitivity}, leading to three key observations.

\begin{enumerate}[leftmargin=*,noitemsep,topsep=0pt]
\item  Tuning parameters can lead to much better performance, compared with using the default parameter configurations. For example, tuning \texttt{scan\_size\_mb}  improves CG performance by 15\%, \textcolor{check}{and tuning \texttt{watermark\_scale\_factor} improves Graph500 performance by 17\%}.  

\item Different workloads exhibit different preferences for parameter configurations. For example, CG achieves optimal performance with \texttt{scan\_size\_mb} set to 8192MB, while Graph500 performs best with \texttt{scan\_size\_mb} set to 256MB. This disparity demands an \textit{adaptive} strategy to adjust parameters based on specific workload requirements. 

\item Tuning certain parameters cannot bring performance benefit. For example, 
tuning \texttt{promote\-\_rate\_limit} and  NUMA scanning-related parameters (except \texttt{scan\_size\_mb}) leads to small performance variance (less than 5\%).
Hence, dynamic tuning of these parameters should not bring large runtime overhead to avoid performance loss.

\end{enumerate}


We do not employ static tuning in our design. With static tuning, a parameter configuration is pre-determined before the workload is launched and does not change throughout workload execution. Static tuning is not practical, because of three reasons. (1) There is no universal static configuration leading to the best performance for all workloads (discussed in Observation 2) and all workload inputs. (2) Static tuning cannot correctly respond to the changing memory access patterns for better performance. (3) Finding a static configuration is challenging, and exhaustive search is time-consuming.

\section{Related Work and Motivation}
\label{sec:motivation}  
{Existing approaches for parameter tuning cannot work well for memory tiering. We classify them  into three categories, each with fundamental limitations.} 

\textcolor{check}{\textbf{Rule-based methods} ~\cite{AMP2020,krish2014hats,baliosian2009rule}, although simple, face difficulty in designing rules that make multiple decisions and account for complex WSs. For example, AMP \cite{AMP2020} uses a rule-based algorithm to 
select the best page migration policy. This policy can \textit{only} be one of three options (LRU, LFU, and random) and is decided based on a \textit{single} WS (page hit ratio).}

\textbf{Heuristic-based methods} ~\cite{doudali2021cori,einziger2018cache, wu2017unimem, ni2023tmc, dulloor2016data_tiering, servat2017automating, vasilakis2020hybrid2, TMO_meta} rely on handcrafted optimizations, which are difficult to be scaled to tune multi-parameters. For example, Cori \cite{doudali2021cori} tunes the page migration frequency based on page reuse distance. 
TMO \cite{TMO_meta} relies on a fixed control policy to decide how much memory to offload from each application to SSD, but the mapping from application profiling to the offload decisions is embedded in handcrafted control logic, and only works for a specific kernel implementation.

\textcolor{check}{\textbf{Machine learning methods} ~\cite{doudali2019kleio, doudali2022coeus,doudali2022cronus, Google_Llama2020, Google_TMTS2023, ren2019archivist, cheng2019optimizing, idt2024chang, doudali2021cori, AMP2020, Google_farmemory2019} are promising but suffer from prohibitive resource requirements and narrow applicability.  For example, Kleio \cite{doudali2019kleio} trains an RNN-based model (i.e., LSTM) \textit{per page} for memory access prediction, resulting in consumption of tens of GBs of memory and 2 hours of training per 100 models.  Google’s warehouse-scale computing (WSC) \cite{Google_farmemory2019} uses a Gaussian Process Bandit model to find the best parameter configuration, requiring one week of WSC’s memory traces to run GP bandit. IDT \cite{idt2024chang} employs RL to adjust the criteria for page demotion, while Cronus~\cite{doudali2022cronus} uses computer vision methods for page access prediction. However, those studies focus on a single aspect of memory management, e.g., access pattern analysis \cite{doudali2019kleio, doudali2022coeus, Google_Llama2020}, page demotion criteria \cite{idt2024chang}, page movement frequency \cite{doudali2021cori}, or page migration policy \cite{AMP2020}, rather than comprehensive parameter tuning.}


\textcolor{check}{Existing solutions also face \textbf{challenges on system integration}. Some solutions (e.g., \cite{idt2024chang, AMP2020, doudali2021cori, Google_TMTS2023}) operate in kernel space, necessitating kernel modifications and recompilation, which is cumbersome in production environments. Some solutions (e.g., \cite{idt2024chang, AMP2020, Google_Llama2020, doudali2022coeus, doudali2022cronus, doudali2019kleio}) focus on optimizing a specific memory management strategy rather than providing a framework applicable to memory tiering solutions with arbitrary configurations.}


\textcolor{check}{In addition, parameter tuning has been explored in various domains~\cite{OtterTune2017, Qtune2019, CDBTune2019, MMDTune2023Ye, MMDTune+2023Sun, MPOD2023, einziger2018cache, CAPES2017, autosys2020, Config-Snob2024}, such as using GP regression and RL for database systems~\cite{Qtune2019, CDBTune2019, MMDTune2023Ye, MMDTune+2023Sun, OtterTune2017}, K-NN for cloud-server energy optimization~\cite{MPOD2023}, and Bayesian optimization for network protocols~\cite{Config-Snob2024}. However, the approaches in those domains do not address the unique challenges faced by memory tiering: the stateful nature of page placement, the need for lightweight runtime decisions, and the requirement to work across various memory tiering solutions without OS modifications.}


\section{Overview}
\begin{figure}[tb!]
	\centering
         \includegraphics[width=0.95\linewidth]{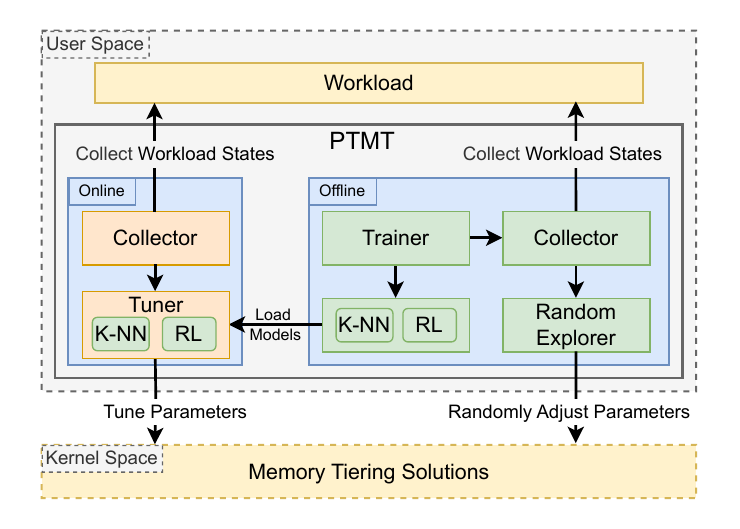}
    \vspace{-10pt}
	\caption{The overview of \name.}
	\centering
	\label{fig:architecture} 
\end{figure}

We introduce \name, an adaptive parameter-tuning framework for memory tiering. The goals of \name are multifold: (1) user-space parameter-tuning that requires no changes to existing memory tiering implementations or the OS, (2) automatic decision-making, and (3) flexibility in customizing tuning objects.


\name defines the workload state (WS), in terms of performance characterization of the workload, and memory tiering state. The memory tiering state is determined based on access traffic to fast and slow memory, such that we can capture the impact of page migration. By formulating the parameter-tuning problem as a stateful problem, \name aims to capture the relationship between the future performance, current WS, and the choice of parameter configurations.

To effectively identify parameter configurations to optimize performance, \name relies on offline WS clustering and online RL, as depicted in Figure~\ref{fig:architecture}. The offline WS clustering is based on our unique observation that WSs under memory tiering  with various parameter configurations can be aggregated into a set of clusters. Given an application, such clusters are built offline, 
using various execution phases and parameter configurations. Data points within each cluster are organized and indexed by a k-nearest neighbor (K-NN) algorithm for fast search. The above offline phase is called the clustering modeling.

During the online phase, \name periodically collects workload information and builds WSs. Given a WS, \name determines whether it falls within any cluster. 
If a match is found, \name identifies the most similar data points within that cluster and then uses the parameter configuration from the data point that has the best performance. 
If the given WS deviates from all clusters, \name uses an RL model to decide the parameter configuration. To accelerate RL convergence, \name pre-trains the RL model offline.  
In general, the clustering model and RL work synergistically to guide the selection of parameter configurations. Figure~\ref{fig:architecture} overviews \name.

\section{Design}

\subsection{Workload State}
\label{sec:state}
The WS consists of (1) workload characteristics and (2) memory tiering state. During application execution, \name samples WSs and decides the configurations of system parameters accordingly. The workload characteristics are represented using performance events. We select two performance events (i.e., LLC hit rate and the second-last-level cache hit) that are most correlated with performance (instruction per cycle or IPC) using Pearson correlation analysis. In our experiments, we observed that using more than two performance events does not yield significant performance improvement. 

\begin{table}[tb!]  
\caption{Metrics to build WS, and the quantification of their correlation to IPC using \textcolor{check}{absolute} correlation coefficient (CC) based on the Pearson correlation analysis.}  

\label{tab:normalized_metrics}  
\centering  
\resizebox{1\columnwidth}{!}{  
\begin{tabular}{|c|c|c|}  
\hline  
\textbf{Metric} & \textbf{Definition} & \textbf{CC} \\   
\hline
\addlinespace[4pt] 
\hline 
$RdR_s$ & Slow-mem read traffic / Total read traffic & 0.82 \\   
\hline  
$WrR_s$ & Slow-mem write traffic / Total write traffic & 0.76 \\   
\hline  
$Tot\_RdR$ & Read traffic / Total memory & 0.98  \\   
\hline  
$L2h$ & L2 cache hit rate & 0.94  \\   
\hline  
$L3h$ & Last level cache hit rate & 0.96  \\   
\hline  
\end{tabular}  
} 

\end{table}

The memory tiering state is designed to represent how effectively fast memory has been utilized by a memory tiering solution. We use slow (or fast)  memory read and write traffic normalized by total memory traffic to represent the memory tiering state. The traffic refers to the total number of bytes loaded/stored within a profiling interval to build the WS. Such a memory tiering state is irrelevant to the memory footprint on fast and slow memories but instead captures the effectiveness of page migration. Ideally, most memory traffic (excluding that related to page migration) should happen on fast memory. In addition, we differentiate between read and write traffic to account for their distinct latency and bandwidth characteristics.

In conclusion, at a given time $t$, the WS  is defined as a vector shown below. Table~\ref{tab:normalized_metrics}  summarizes WS along with the correlation to IPC. 

\begin{equation}
\label{eq:ws}
    WS(t) = [\, L2h, \;\; L3h, \;\; RdR_s, \;\; WrR_s, \;\; Tot\_RdR \,]
\end{equation}

\subsection{Clustering of Workload States}
\textbf{Definition of the search space.} The search space is a collection of data points where each data point is a vector consisting of $WS(t)$ at a time $t$, the parameter configurations to be applied $Params(t)$, and the aftermath performance $IPC(t')$ after applying the configurations. The search space is large. Given an application, we can use various application inputs and measure $WS$ throughout application execution to sample the search space. 

\begin{figure}[tb!]
	\centering
         \includegraphics[width=1.0\linewidth]{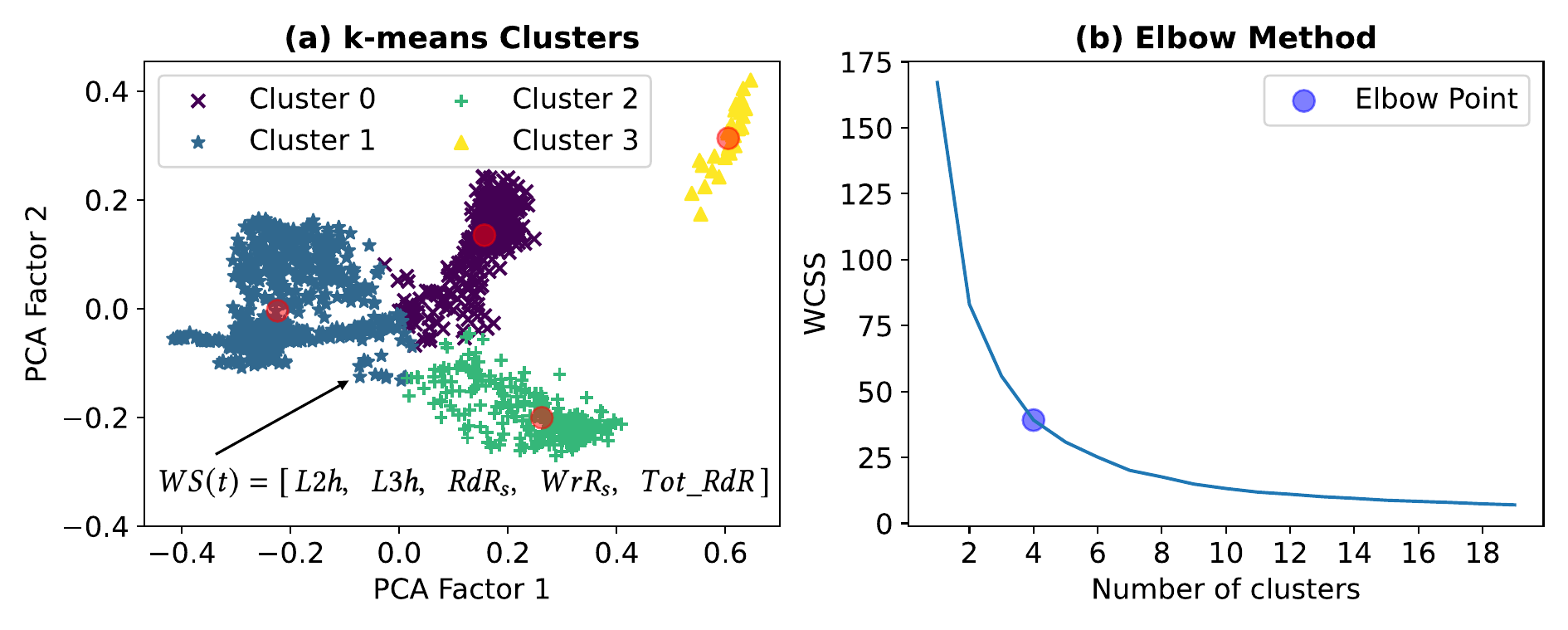}
    \vspace{-20pt}
	\caption{k-means clustering of \textit{WSs} in the benchmark NPB-LU. Each dot represents a WS. Red dots are the cluster centroids. Different colors indicate different clusters.}
	\centering
	\label{fig:kmeans_clusters} 
\end{figure}

\textbf{Clustering WSs.} We observe that WSs in data points in the search space can be clustered.  
Figure~\ref{fig:kmeans_clusters}.a shows the distribution of WSs for the benchmark NPB-LU~\cite{NASA_NPB} (a Lower-Upper Gauss-Seidel solver) as an example. We execute LU 100 times with AutoNUMA and various parameter configurations,  using the same input (Class E), and collect 6K WSs.


We run the k-means~\cite{kmeans} to cluster those WSs. To determine the appropriate number of clusters (i.e., $k$), we employ the elbow method~\cite{cluster_elbow}, a standard technique to identify the point where adding more clusters provides diminishing returns in terms of the reduction of a metric, Within-Cluster Sum of Squares (WCSS). Using the elbow, we ensure that our clustering model captures sufficient granularity to distinguish between different WSs without over-fitting. Figure~\ref{fig:kmeans_clusters}.b indicates that ``4'' is the optimal number of clusters for NPB-LU. 

\begin{figure}[tb!]
	\centering
         \includegraphics[width=1.0\linewidth]{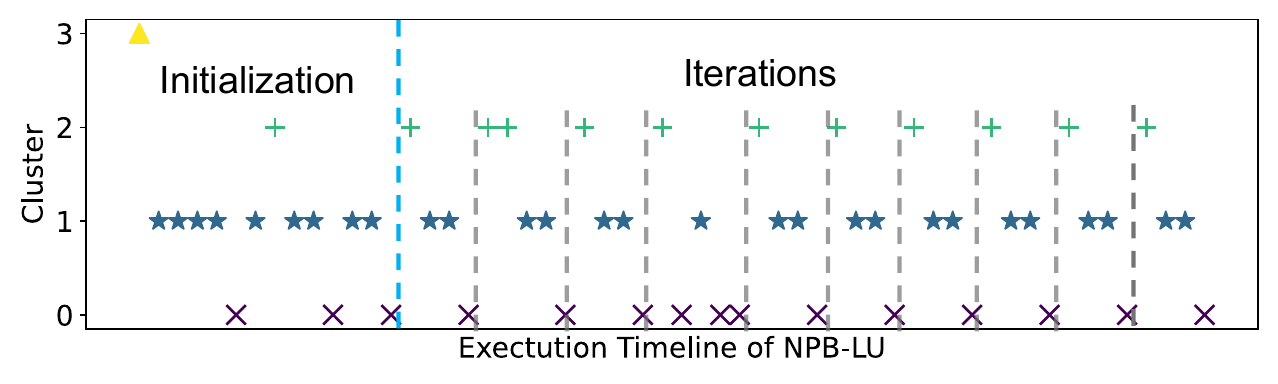}
    \vspace{-20pt}
	\caption{Map WSs and clusters in Figure~\ref{fig:kmeans_clusters} to execution phases. Each marker in Figure~\ref{fig:cluster_time} represents about 100 points.}
	\centering
	\label{fig:cluster_time} 
\end{figure}

We find that the WS clusters correspond to execution phases, even though WSs are collected with various parameter configurations. See Figure~\ref{fig:cluster_time}. The figure maps WSs and the corresponding clusters to the execution phases. For example, the marker ``$\times$'' corresponds to a data copy phase; the marker ``$+$''  corresponds to the computation of the regular-sparse, block lower triangular solution.

The above results indicate that given an input to the application, WSs collected from the same execution phase, even if they are collected using different parameter configurations and have diverse memory-tiering states, tend to be clustered together. WSs collected from different execution phases have difference in \textit{both} workload characteristics and memory tiering state, tend to go to different clusters. Also, using more application inputs to collect WSs, we add more clusters. In general, WSs are not uniformly distributed in the space.

The reason for the above clustering results is that memory tiering is a fine-tuning process to distribute memory access traffic between fast/slow memories and avoid frequent page migration~\cite{meta_tpp}. As a result, the memory tiering state for an execution phase does not change significantly across different parameter configurations, compared to the change in workload characteristics across execution phases.

\textbf{Application-specific clustering.} For each application, we build a clustering model. We chose application-specific clustering rather than general clustering across all applications to improve the effectiveness of parameter tuning. Our evaluation shows that using application-specific clustering outperforms the general clustering by 20\% across four memory tiering solutions (see Sec.~\ref{sec:eval_rl}).  


For a single application, building application-specific clustering requires  $11\times$ less time than building a general clustering model (see Sec. ~\ref{sec:eval_rl}). 

The application-specific clustering is feasible because it aligns with the common deployment practices of major public-cloud providers for HPC and AI. Those vendors focus on a narrow set of applications in a pre-defined domain. This practice is especially common in the business of HPC- and AI-centric SaaS  \cite{rescale-hpc-saas,simscale-saas}. For example, Amazon Web Services (AWS)  provides HPC clouds \cite{aws-hpc-hcls,aws-hpc-cfd}, focusing on individual domains (e.g., healthcare and life science). In this type of customized environment, a specific modeling and simulation workload (e.g., computational fluid dynamics (CFD), pharmacokinetics, clinical trial simulation, or systems biology) is frequently launched. Google Cloud Platform (GCP) provides similar services. For example, GCP provides HPC blueprints \cite{gcp-cae} for computer-aided engineering (CAE) that target standard solvers such as Ansys Fluent, Siemens Simcenter STAR-CCM+, and OpenFOAM. We have similar observations in AI-centric services where cloud vendors offer catalogs of ready-to-use AI models and pre-built training and inference pipelines \cite{azure-model-catalog,vertex-model-garden}.

\textbf{Improving similarity match within the cluster.}
To narrow down the search space in a cluster, given a WS, we employ a feature-weighted K-nearest neighbors algorithm (K-NN)~\cite{feature_weighted_knn} to find data points most similar to the given WS. The similarity is defined in terms of workload characteristics and memory tiering state.

The feature-weighted K-NN assigns different weights to different features (i.e., the metrics listed in Table \ref{tab:normalized_metrics}), and calculates the distance between two data points based on the weighted Euclidean distance. To determine appropriate weights for features, 
we construct a decision tree using data points in the cluster, with IPC specified as the decision target. In this decision tree, the internal nodes represent features, and the leaves represent IPC predictions. After the construction, a feature importance value is calculated based on the feature's contribution to IPC prediction. The importance value is used as the feature weight for K-NN. 
We use the above approach, because it differentiates features according to their importance to performance, hence allowing us to find the most similar data points. 




\subsection{Reinforcement Learning}

We employ RL to tune parameter configurations at runtime.

\subsubsection{Background}
RL is a type of ML that makes decisions to achieve the best outcome by interacting with the \textit{environment} and learning through an \textit{agent}. Specifically, given an environment $State(t)$ at a time step $t$, the agent takes $Action(t)$, which causes a transition to the next $State(t+1)$. The agent then receives $Reward(t+1)$ for moving from $State(t)$ to $State(t+1)$ based on $Action(t)$. An \textit{episode} in RL is a complete cycle of interactions that starts in an initial state and ends when a terminal state is reached. The policy $\pi$ guides the action the agent performs in a specific state. RL aims to find the optimal policy $\pi^{*}$ that yields the highest cumulative reward to achieve an optimal outcome. We employ Proximal Policy Optimization (PPO) \cite{RL_PPO}, a widely used and robust RL  algorithm, to find the optimal policy $\pi^{*}$. 

\textbf{Why RL?} RL can benefit parameter tuning for memory tiering for three reasons. First, the RL workflow aligns naturally with the stateful parameter tuning process. 
In memory tiering, the selection of parameter configurations \textcolor{check}{affects} the way WS transits from one state to another, ultimately affecting system performance. 
RL excels in such a context by making decisions that influence state transitions, thus optimizing the  outcome over time. 

Second, RL's ability to continuously learn and adapt to changing environments is crucial to respond to dynamics in WS. RL can update its decision-making processes in real time, allowing it to make better choices as WS evolves. 


Third, RL naturally handles scenarios with application co-runs. Co-running applications create a combinatorially large search space, making it infeasible to collect sufficient data points within a reasonable time budget. By treating a set of co-running applications as a single aggregate workload, RL can learn the evolving mixed WS and continuously refine its policy at runtime.

The RL model in \name is a lightweight three-layer Multi-Layer Perceptron (see Sec.~\ref{sec:overhead} for overhead analysis). 
Hence, RL effectively addresses the limitation discussed in Sec.~\ref{sec:motivation}.

\subsubsection{RL Formulation}
We define the key elements of RL. 

\textit{Optimization goal} is to maximize application performance by finding a policy $\pi^*$ over application's lifetime $T$.
\begin{equation}
    \pi^{*} = \arg\max_{\pi} \frac{1}{T} \sum_{t=1}^{T} IPC(t)
\end{equation}

\textit{State} is the current environment information and RL performs an action upon it. Equation~\ref{eq:ws} defines the state. 


\textit{Action} represents the parameter configurations that \name applies to different WSs to optimize performance. The action space consists of possible parameter configurations. At any time $t$, the action is a specific set of parameter configurations chosen from the action space, described as follows.
\begin{equation}
    \begin{aligned}
        Action(t) = \{(P_1(t), \,\, P_2(t), \, \dots, \, P_n(t))\}
    \end{aligned}
\end{equation}
where $P_1(t), \, P_2(t), \, \dots, \, P_n(t)$ represent individual parameter tuning at $t$, and $n$ is the number of parameters to tune.


\textit{Reward} reflects the effect of the previous action performed on the previous state. As we aim to improve performance, we directly measure the reward based on IPC. We normalize IPC such that it falls in [-1, 1]. Otherwise, the reward received by the agent is always a positive value, failing to penalize the suboptimal action. 
\begin{equation}
    \begin{aligned}
        Reward(t) &= 2 \left( \frac{IPC(t) - IPC_{min}}{IPC_{max} - IPC_{min}} \right) - 1
    \end{aligned}
\end{equation}
where $IPC_{min}$ and $IPC_{max}$ represent the lower and upper bounds of IPC collected offline.

\subsubsection{RL Model Structure and Hyper-Parameters} 

The RL model consists of three layers. The first layer comprises five neurons, corresponding to the five metrics listed in Table~\ref{tab:normalized_metrics}. The second layer is a fully connected hidden layer with 64 neurons. The third layer varies in size, depending on the number of parameters to tune. 

For hyper-parameters in RL, \textcolor{check}{according to the sensitivity study in Sec.  \ref{sec:eval_sensitivity},} we set the discount factor to 0.9 and the learning rate to 0.01. In PPO, as Stable-Baselines3 (SB3) \cite{stable-baselines3}, actions are explored based on probability distributions predicted for each action by RL, without the use of an explicit exploration rate. We set the rollout buffer size (the number of steps per policy update) to 4. 

\subsubsection{RL Convergence Acceleration}
The RL model convergence in our scenario can be slow because of the large exploration space. Assuming that $S$ is the number of possible WSs and $P$ is the number of parameter configurations in the action space, the capacity of the exploration space is $S^P$, which is large. Directly applying RL for online tuning degrades application performance because the agent may require numerous episodes to perform extensive explorations.

To accelerate model convergence and provide a solid baseline for online tuning, we pre-train the RL model using Behavioral Cloning (BC)~\cite{behavioral_cloning}. BC is an ML technique. Using BC, an agent learns to perform tasks by mimicking the behavior of an expert. To generate the expert dataset required for BC, we reuse our clustering model. Specifically, for each $WS$ collected offline, we identify the cluster to which it belongs and then employ K-NN within that cluster to find the $k$ most similar data points. Among these data points, we search for the one with the best IPC and label the corresponding  $Params$ for the given WS. After processing all WSs, we obtain optimal WS-$Params$ pairs. Those WS-$Params$ pairs serve as the expert dataset in BC. Using this expert dataset, the RL agent can quickly learn a policy by directly mapping optimal $Params$ to WS without extensive exploration.

\subsubsection{Hybrid Method} 
\label{sec:hybrid}
By default \name uses the clustering model, which is effective when 
WS measured at runtime is close to one of the centroids and falls into one of the clusters. However, when WS deviates substantially from all centroids, the clustering model cannot lead to optimal parameter configurations. In such a case, the RL model becomes more applicable. The switch between the clustering model and RL happens automatically in \name.  

To decide whether WS is outside all clusters, for each cluster, \name calculates the mean ($\mu$) and standard deviation ($\sigma$) of the distance of all data points from the centroid. If the distance between $WS$ and the centroid is larger than a threshold ($\mu + 3 \sigma$), WS is considered to be outside the cluster. The threshold is statistically determined based on the assumption that the distance of the data points from the centroid follows a normal distribution, and hence 99.7\% data points have a distance less than $\mu + 3 \sigma$ ~\cite{normal_dist}. WS with a distance greater than this threshold is considered an outlier. 

\subsection{More Implementation Details}
We implement \name as a user-level framework without requiring OS or application modifications, making it applicable to production environments.
We 
verified the compatibility of \name with Linux kernel 5.13, 6.1, and 6.3. 

\textcolor{check}{\textbf{Workload states collection.}  \name employs Intel Performance Counter Monitor (PCM) \cite{intel_pcm} and AMD Instruction Based Sampling (IBS) \cite{amd_ibs} to collect WSs. 
To build the performance database, for each workload, we repeatedly run it with multiple application inputs, collecting WS and IPC, and randomly setting parameter configurations at each tuning interval until 3K data points are gathered.}

\textcolor{check}{\textbf{Models.} For the clustering model, \name uses k-means algorithm to build clusters and builds K-NN within each cluster, supported by Python’s \texttt{scikit-learn}. For RL model, \name uses PPO from Stable-Baselines3 (SB3) \cite{stable-baselines3} as the policy network in RL. Before online tuning, \name uses BC to pre-train PPO as the initial policy in RL.}




\textcolor{check}{\textbf{Parameter adjustment.} For kernel-based memory tiering solutions (AutoNUMA, Colloid, and TPP), \name tunes parameters by writing values to files under \texttt{/sys/\allowbreak kernel/\allowbreak debug} and \texttt{/proc/\allowbreak sys}. For the user-space UPM, \name uses its client-server interface via command-line instructions.}

\textbf{Overhead control.} To minimize CPU resource consumption, \name uses a single helper thread for WS collection and parameter tuning. 
\textcolor{check}{Both the RL training and inference processes also operate on this thread.}
We analyze the overhead in Sec.~\ref{sec:overhead}.

\section{Evaluation}
\label{sec:eva}

\subsection{Evaluation Methodology}
\label{sec:eval_method}
\textbf{Evaluation platform.} We evaluate \name on a dual-socket machine equipped with Intel Xeon Gold 6252 @2.10 GHz processors (24 cores per socket). Each socket has 6×16GB DDR4 DRAM as fast memory, and 6×128GB Intel Optane DCPMM as slow memory. To study the impact of page migration with memory tiering and avoid cross-socket NUMA effects, we conduct all experiments on a single socket, similar to prior works ~\cite{Memtis2023, meta_tpp, atc24_hm}. We use \texttt{GRUB mmap} to limit fast memory capacity to 34GB, such that the workload memory is allocated on both fast memory and slow memory.  

\textbf{Memory tiering settings. } For AutoNUMA, \textcolor{check}{Colloid,} and TPP, we enable page promotion by setting \texttt{/proc/sys} \texttt{/kernel/numa\_balancing} to 2, and enable page demotion by setting \texttt{/sys/kernel/mm/numa/demotion\_enabled} to 1. 

\textbf{Benchmarks. } We use seven memory-intensive benchmarks summarized in Table \ref{tab:benchmarks}. The benchmarks consist of a graph processing benchmark (Graph500 \cite{graph500}), an in-memory database engine (Silo~\cite{silo}), a machine learning benchmark (Liblinear~\cite{liblinear}), and four high performance computing benchmarks (LU, SP, BT, and FT) from NPB~\cite{nas}.

During the offline phase, we collect WSs to generate the clustering model and RL pre-trained model using some input problems  (named WSS (a)). During the online phase, we assess the effectiveness of \name by measuring the performance of each benchmark under two scenarios: (1) using similar input problems as in the offline phase (WSS (a)), such that WS collected online falls into the clusters built offline; (2) using largely different input problems, such that WS may randomly fall out of the clusters (named WSS (b)).   


\textbf{Baselines.} We compare \name performance against that of using the default parameter configurations. Each benchmark is executed using 24 threads, and execution time is measured to assess performance. 
Besides \name and the four memory tiering solutions, we evaluate \texttt{NoBalance} in Linux by disabling NUMA page migration, which represents static page placement. 
We also evaluate IDT~\cite{idt2024chang}, a state-of-the-art solution that uses an RL-based policy to efficiently demote cold pages to
slow memory. IDT uses RL to tune page demotion criterion \texttt{age\_thres}, which is a threshold determining whether a page is ``aged'' enough for demotion.
Specifically, IDT uses a function of  each memory region's age as RL's state and uses the performance (inversely proportional to the slow memory access frequency) as RL's optimization goal to determine the appropriate \texttt{age\_thres}.
IDT is not application-specific, but pre-trained using a general micro-benchmark GUPS \cite{gups} emulating three common access patterns: uniform random, static hot set (where 90\% of accesses happened to a static hot-memory region), and dynamic hot set (where the hot set changes every 150 seconds).

\textbf{Parameter selection for tuning.} Parameters are selected based on performance sensitivity study (some results are shown in Sec. \ref{sec:sys_perf_senstivity}). A parameter is selected if at least one benchmark's performance is sensitive to its configuration. Table \ref{tab:parameter_selection} summarizes the parameters for tuning.

\begin{table}[t!]
\caption{Benchmarks for evaluation}
\label{tab:benchmarks}
\centering
\resizebox{1\columnwidth}{!}{
\begin{tabular}{|c|c|c|c|c|}
\hline
\textbf{Benchmark} & \textbf{Description} & \textbf{WSS (a)} & \textbf{WSS (b)} \\ 
\hline
\addlinespace[4pt] 
\hline
LU & Lower-Upper Gauss-Seidel solver & 134GB & 58GB  \\ \hline
SP & Scalar Penta-diagonal solver & 174GB  & 74GB  \\ \hline
BT &  Block Tri-diagonal solver & 166GB & 71GB  \\ \hline
FT & Discrete 3D fast Fourier Transform & 80GB & 40GB  \\ \hline
Graph500 & Graph generation and traversal & 64GB & 46GB  \\ \hline
Liblinear & Large-scale linear classification & 69GB & 59GB  \\ \hline
Silo & In-memory database engine & 115GB & 145GB  \\  \hline

\hline
\end{tabular}
}
\end{table}





 


\begin{table}[t!]
\centering
\caption{Parameters to tune in memory tiering}
\label{tab:parameter_selection}
\footnotesize
\begin{tabular}{|c|c|}
\hline
\textbf{Memory Tiering Solution} & \textbf{Parameters} \\ 
\hline
\addlinespace[4pt] 
\hline

\multirow{2}{*}{AutoNUMA}
  & \texttt{scan\_size\_mb}          \\ 
  & \texttt{watermark\_scale\_factor}\\ \hline

\multirow{2}{*}{\textcolor{check}{Colloid}}
  & \texttt{scan\_size\_mb}          \\
  & \texttt{\textcolor{check}{watermark\_scale\_factor}}  \\ \hline

\multirow{2}{*}{TPP}
  & \multirow{2}{*}{\texttt{demote\_scale\_factor}} \\ 
  &                                               \\ \hline

\multirow{3}{*}{UPM}
  & \texttt{hot\_threshold}               \\
  & \texttt{sample\_period}               \\
  & \texttt{page\_migration\_interval}    \\ \hline

\end{tabular}
\end{table}
\vspace{-10pt}

\subsection{Overall Performance}
We evaluate two scenarios: (1) assessing the effectiveness of the clustering model using WSS (a), and (2) assessing \name's adaptiveness using WSS (b).




\begin{figure*}[tb!]
	\centering
         \includegraphics[width=1\linewidth]{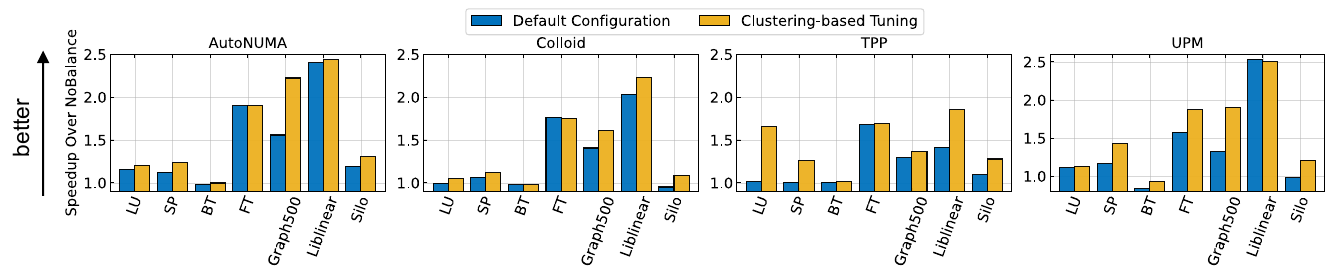}
    \vspace{-20pt}
	\caption{WSS (a): Performance speedup over \texttt{NoBalance} performance.}
	\centering
	\label{fig:eval_overall_wss_a}

\end{figure*}

\begin{figure*}[tb!]
	\centering
         \includegraphics[width=1\linewidth]{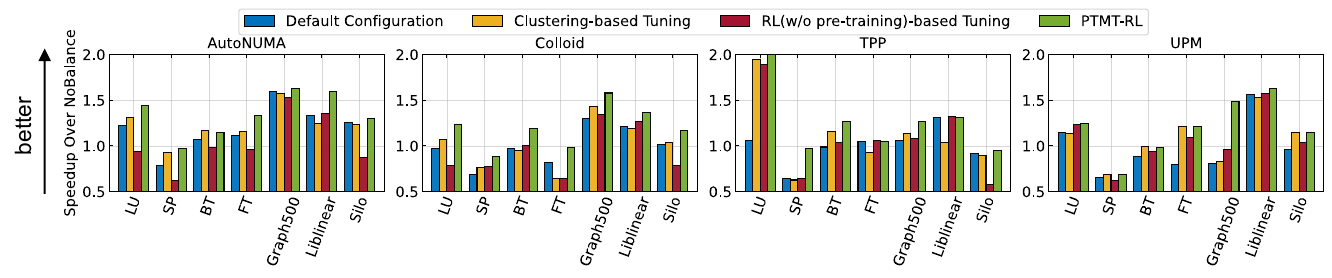}
    \vspace{-20pt}
	\caption{WSS (b): Performance speedup over \texttt{NoBalance} performance.}
	\centering
	\label{fig:eval_overall_wss_b} 
\end{figure*}

\begin{figure*}[tb!]
	\centering
         \includegraphics[width=1\linewidth]{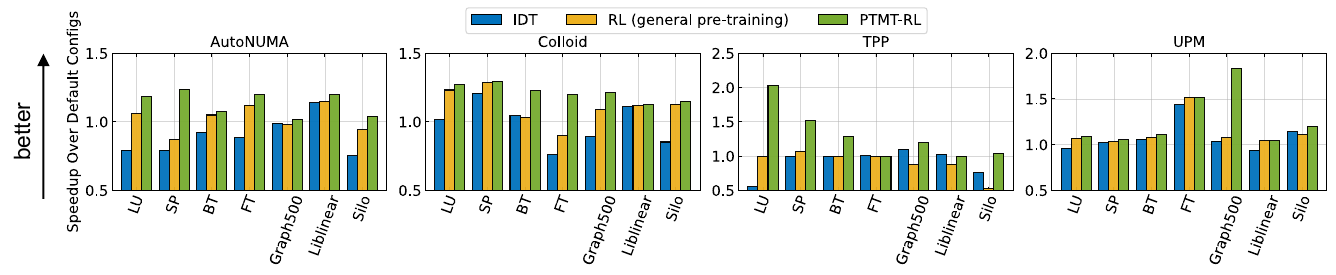}
    \vspace{-20pt}
	\caption{WSS (b): Evaluation of application-specific RL. The performance speedup is measured over the default configuration.}
	\centering
	\label{fig:eval_rl_wss_b} 
\end{figure*}

\textbf{WSS (a).} 
The clustering-based tuning without RL takes effect, as shown in Figure \ref{fig:eval_overall_wss_a}. 
\name improves the performance of AutoNUMA, \textcolor{check}{Colloid,} TPP, and UPM by \textcolor{check}{10\%, 7\%, 20\%, and 17\%} on average, respectively, compared to their default configurations. Specifically, while the default outperforms the NoBalance baseline by an average of \textcolor{check}{47\%, 31\%, 21\%, and 36\%} on AutoNUMA, \textcolor{check}{Colloid,} TPP, and UPM, respectively, the clustering-based tuning outperforms NoBalance by \textcolor{check}{62\%, 40\%, 44\%, and 57\%} on the same memory tiering solutions. This demonstrates the effectiveness of the clustering model. 


We observe that some workloads are very sensitive to parameter tuning in specific memory tiering solutions. For example, compared to the default, the clustering‑based tuning outperforms by 10\% on SP, 43\% on Graph500, and 10\% on Silo in AutoNUMA and \textcolor{check}{by 14\% on Graph500, 10\% on Liblinear, and 14\% on Silo} in \textcolor{check}{Colloid}. However, the clustering‑based tuning achieves little performance improvement for BT and FT in AutoNUMA and \textcolor{check}{Colloid}.


The best performance improvements are achieved on  Graph-500 when using \name  with AutoNUMA and \textcolor{check}{Colloid}. Further analysis reveals that both AutoNUMA and \textcolor{check}{Colloid} employ NUMA scanning and hint fault latency for memory profiling and hotness detection, and incorporate flexible page migration mechanisms. 
\textcolor{check}{The memory profiling speed is controlled via \texttt{scan\_size\_mb}, and the page demotion aggressiveness is controlled via  \texttt{watermark\_scale\_factor}. This flexibility allows \name to fine-tune memory profiling for quicker detection of changes in hot page set and to adjust the proactiveness of page demotion in response. This adaptability is suited to Graph500’s frequent changes in the hot page set.}

In all cases (except for Liblinear with UPM), \name does not cause performance loss compared to the default configurations. Liblinear with UPM and \name is the only exception. Further analysis reveals that, unlike other solutions, UPM's page demotion appears on the critical path of page promotion, slowing down the promotion of hot pages. \name amplifies this issue, resulting in reduced performance.



\textbf{WSS (b).} Figure \ref{fig:eval_overall_wss_b} shows that with RL, \name outperforms the default configurations by \textcolor{check}{14\%, 21\%, 30\%, and 26\%} on average for AutoNUMA, \textcolor{check}{Colloid}, TPP, and UPM, respectively. 
\name automatically enables RL model to tune parameters when WSS changed. 
These results demonstrate \name effectiveness in tuning parameters across a diverse set of WSs.


Compared to the default configurations, \name achieves  performance gain by 103\% for LU with TPP, while other memory tiering solutions do not have a large gain. 
That substantial gain is due to TPP’s more aggressive demotion mechanism, which employs an asynchronous background process. LU's hot regions are distributed across fast and slow memories, hence calling for an effective demotion mechanism to accommodate emerging hot pages. By tuning  \texttt{demote\_scale\_} \texttt{factor} in TPP, \name optimizes demotion aggressiveness, improving page placement and overall performance.


\name's clustering-based tuning degrades Liblinear's performance across all memory tiering solutions. This occurs because WSS (b) has approximately 80\% of its WSs falling outside the clusters (formed using WSS (a)), indicating that the clustering-based tuning lacks the flexibility needed to adapt to the diverse WSs in a workload.

\subsection{Effectiveness of RL}
\label{sec:eval_rl}



\textbf{Adaptiveness of RL.} We compare the clustering-based,  RL-based \textcolor{check}{(PTMT-RL)}, and  RL (without pre-training)-based tuning using WSS (b). Figure~\ref{fig:eval_overall_wss_b} shows that the RL-based tuning outperforms the clustering-based tuning by \textcolor{check}{9\%, 21\%, 19\%, and 14\%} on average for AutoNUMA, \textcolor{check}{Colloid}, TPP, and UPM, respectively, indicating that RL is more effective in handling diverse WSs.


RL with pre-training outperforms without pre-training by \textcolor{check}{34\%, 31\%, 24\%, and 14\%} on average for AutoNUMA, \textcolor{check}{Colloid,} TPP, and UPM, respectively. This improvement is because the pre-training phase provides RL with prior knowledge, reducing the number of iterations needed to explore parameter configurations and develop an optimal policy. 

\begin{figure}[tb!]
	\centering
    \includegraphics[width=1\linewidth]{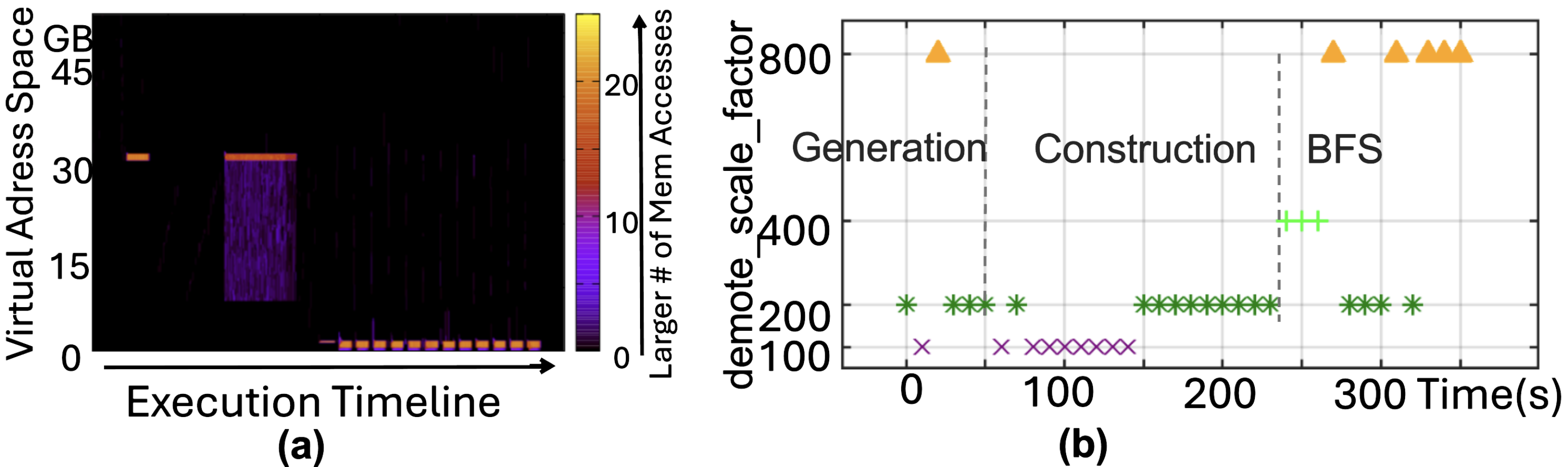}
    \vspace{-20pt}
	\caption{(a) Memory access heatmap of Graph500. (b) Tuning \texttt{demote\_scale\_factor}  using \name. The markers in the figure represent parameter configurations.}
	\centering
	\label{fig:access_pattern_wss_b} 
\end{figure}

\begin{figure}[tb!]
	\centering         \includegraphics[width=1\linewidth]{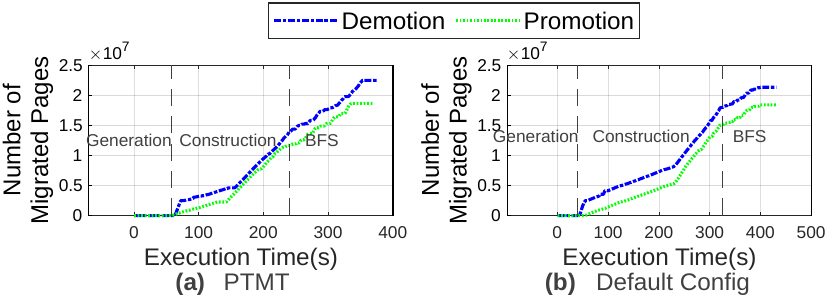}
 \vspace{-15pt}
 \caption{Page migration when tuning Graph500.}
	\centering
\label{fig:tpp_graph500_s_tune_page_migration} 
\end{figure}

\textbf{Effectiveness of application-specific RL.} 
We compare with IDT. 
Since IDT is only used to tune a specific parameter for a specific memory tiering solution (see Sec. \ref{sec:eval_method}),
we improve IDT in three ways (see our repo~\cite{ptmt}): 
(1) enabling IDT to tune multiple system parameters rather than a single parameter, (2) incorporating WSs as the RL state representation, and (3) making the decision epoch configurable as a tuning period. We pre-train four separate models, each tailored to one of the four memory tiering solutions.
To train each pre-trained model, we repeatedly run GUPS with three distinct memory access patterns and five different hot-region distributions until adding more iterations no longer significantly reduces the loss of RL's value function, indicating that the model's estimates of state values are stable. This general pre-training process leads to $11\times$ more execution iterations than those using application-specific models in \name. 

Following IDT's approach to build a general pre-trained model, we pre-train RL (named ``RL (general pre-training)'') and compare it with RL that uses the application-specific pre-training (named ``\name-RL''). 
To build the general model offline, we run GUPS with the same memory access patterns and iteration counts as IDT to construct a performance database for RL (general pre‑training). After generating the expert dataset from this performance database, we apply Behavioral Cloning to pre-train PPO as the initial policy in RL. To optimize the performance of this RL (general pre-training) for online tuning, we use sensitivity study to decide its hyper-parameters.

Figure \ref{fig:eval_rl_wss_b} shows the results. Compared to the default configuration, IDT improves performance by \textcolor{check}{-10\%, -1\%, -8\%, and 9\%} for AutoNUMA, \textcolor{check}{Colloid}, TPP, and UPM, respectively. RL with general pre-trained improves performance by \textcolor{check}{2\%, 11\%, -10\%, and 13\%}. These results suggest that a general RL model has difficulty capturing application-specific optimization, even if it collects various WSs from a general micro-benchmark. In contrast, the application-specific RL significantly improves performance: \name-RL outperforms IDT by \textcolor{check}{29\%, 26\%, 56\%, and 17\%} for AutoNUMA, \textcolor{check}{Colloid}, TPP, and UPM, respectively, and outperforms RL (general pre-training) by \textcolor{check}{12\%, 10\%, 46\%, and 12\%} respectively. 

\subsection{Analysis of Dynamic Parameter Tuning}




We look deep into how the parameters are tuned at runtime by \name. We use Graph500 (WSS (b)) with TPP as an example, shown in Figure \ref{fig:access_pattern_wss_b}.b. We then provide an in-depth comparison of parameter tuning and page migration between \name and the default configuration (i.e., \texttt{demote\_scale\_\allowbreak 
 factor} = 200), shown in Figure \ref{fig:tpp_graph500_s_tune_page_migration}. Figure \ref{fig:access_pattern_wss_b}.a shows memory access heatmap with execution time. 


Graph500 has three execution phases, as shown in Figure \ref{fig:access_pattern_wss_b}.b. In the generation phase where edge lists are generated, \name uses the default configuration and there is minor page migration. This phase is characterized with limited page reuse, and hence lack hot pages. \name recognizes this and keeps the parameter configuration.



In the graph construction phase where the graph structure is built from the edge lists, \name uses a smaller value (``100'') for the parameter. This phase has a rather large hot-page set, inevitably spilled to slow memory. Hence, the memory tiering should avoid page thrashing between memory tiers~\cite{meta_tpp}. Using the smaller value, TPP is able to reduce unnecessary page demotion by 15\% and page promotion by 20\%, compared with using the default configuration. Thanks to \name, this phase improves performance by 32\% (see Figure \ref{fig:tpp_graph500_s_tune_page_migration}).


In the BFS phase, the hot region becomes smaller and stable (see Figure~\ref{fig:access_pattern_wss_b}.a) because Graph500 uses a stochastic Kronecker graph generator to create a power-law graph  where a few vertices hold most edges. \name uses a larger value (400 or 800), enabling more aggressive page migration. Figure \ref{fig:tpp_graph500_s_tune_page_migration} shows 2.1$\times$ more page demotion and 2.0$\times$ more page promotion, compared to using the default configuration.

In conclusion, \name improves the performance of Graph500 by 20\%, and \name responds correctly to WSs throughout application execution.

\begin{figure*}[t!]
	\centering         \includegraphics[width=1\linewidth]{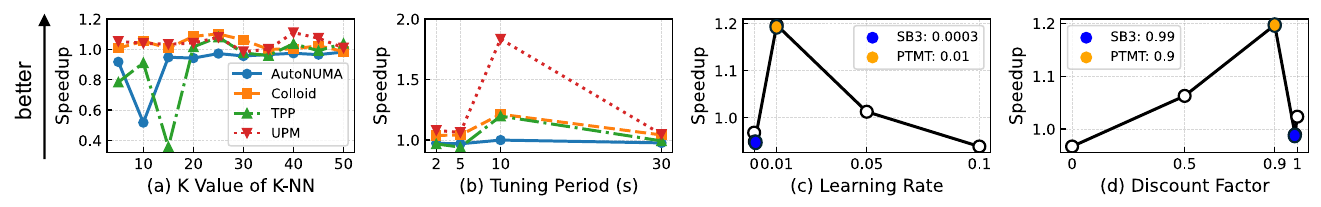}
    \vspace{-20pt}
	\caption{Sensitivity study to hyper-parameters. Performance speedup is over that of the default configuration.}
	\centering
	\label{fig:hyperparameter_sensitivity} 
\end{figure*}

\begin{figure}[!t]
	\centering         \includegraphics[width=1\linewidth]{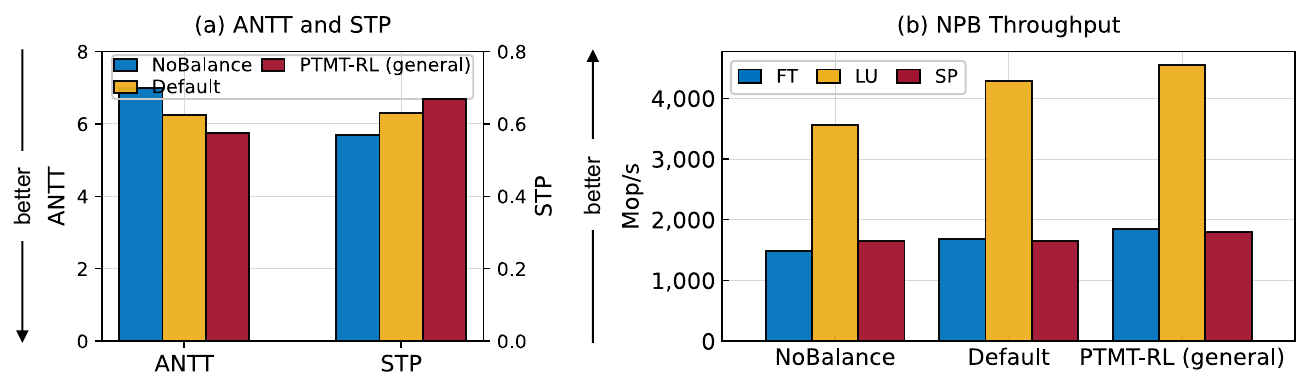}
    \vspace{-15pt}
	\caption{Overall ANTT-STP and per-application throughput (Mop/s) for application co-run.}
	\centering
	\label{fig:multi_apps} 
\end{figure}

\subsection{\name with Application Co-run}
We launch FT, LU, and SP  simultaneously and compare NoBalance, the default configuration, and \name‑RL (using general pre-training on GUPS) under AutoNUMA. Because \name uses system‑wide performance metrics to represent WSs and tune system‑wide parameters accordingly, it can treat co‑running applications as a single aggregate workload by collecting global WSs for the entire system. 
In the case of application co-run, pre-training RL for each combination of co-running applications is not scalable. Therefore, we use the general pre-training method (see Sec. \ref{sec:eval_rl}). In particular, we use the general pre-trained RL model and tune the model parameters when co-running FT, LU, and SP (WSS(b)) based on the collection of global WSs.

We use two metrics, System Throughput (STP) and Average Normalized Turnaround Time (ANTT) \cite{eyerman2008system, wu2015enabling}. STP reflects system-wide throughput (the larger is better), while ANTT captures the average slow-down experienced when multiple applications share server resources (the smaller is better).
We also report the per-application throughput (Mop/s, millions of operations per second), provided by NPB.

Figure \ref{fig:multi_apps} shows the results: compared with NoBalance and the default, \name-RL (using general pre-training) reduces ANTT by 18\% and 8\%,  increases STP by 18\% and 8\%, and boosts application throughput by 21\% and 8\%, respectively. This indicates that, with general pre-training, the PTMT solution is applicable to application co-run cases.




\subsection{Sensitivity Study}
\label{sec:eval_sensitivity}

We use Graph500 (WSS (b)) as an example. We see similar results from the sensitivity study for other benchmarks. 


\textbf{K value in K-NN.} We vary K using the cluster modeling. Figure \ref{fig:hyperparameter_sensitivity}.a shows that when K < 25, the performance degrades due to a limited number of nearest WSs to find the best performance. But when K >  25, the performance variance is less than 10\%. We choose K$=25$ for \name.


\textbf{Tuning period.} We vary the tuning period from 2 to 30 seconds. Figure \ref{fig:hyperparameter_sensitivity}.b shows that short (2-5s) and long (30s) tuning periods bring trivial performance improvement. A short period fails to give the workload to respond to parameter tuning, while a long period misses opportunities to tune parameters. We choose 10 seconds for \name.

\textbf{RL hyperparameters.}
We study two key RL hyperparameters: the learning rate and discount factor. 
The learning rate dictates how quickly the RL updates its network weights. {Figure \ref{fig:hyperparameter_sensitivity}.c shows that a learning rate of 0.01 outperforms  the default setting in SB3~\cite{stable-baselines3} (0.0003) and a higher rate (0.1). Because a lower rate reduces the model's adaptability to dynamic environments, while a higher rate increases computational overhead of online training due to frequent weight updates. We choose 0.01 for \name.




The discount factor determines the weighting of past and immediate rewards. Figure \ref{fig:hyperparameter_sensitivity}.d reveals that the discount factor of 0.9 performs better than 0 or 0.99 (the default setting in SB3). This suggests that focusing solely on immediate rewards or placing excessive emphasis on past rewards leads to suboptimal performance. ``0.9'' balances the tradeoff. 

\begin{table}[tb!]  
    \centering 
    \caption{Execution time for \name components} 
    \small
    \label{tab:overhead}
    \begin{tabular}{|c|c||c|c|}  
       \hline
        \textbf{Component}                & Time (ms) & \textbf{Component}                &Time (ms) \\ 
\hline
\addlinespace[4pt] 
\hline
        
        WS Collection            & 21                   & Clustering Query         & 24                \\ 
        \hline
        
        RL Inference             & 1                    & Model Switch             & 1                    \\ 
        \hline

        Config Setting           & 9 - 18               & RL Policy Update         & 84                   \\ 
        
       \hline 
    \end{tabular} 
\end{table}

\subsection{Overhead Analysis}
\label{sec:overhead}

\textcolor{check}{For storage overhead, the offline-generated clustering model (built from $\sim3,000$ data points) requires about 3 MB, and the pre‑trained RL model requires 70 KB. During online tuning, the transient $WS(t)$ consumes only 96 bytes (12 FP numbers).}

For computation overhead, we measure the time for various components in \name, as summarized in Table~\ref{tab:overhead}. The overall performance overhead of \name is less than 1.0\% when the tuning period is set to 10 seconds. 
Collecting $3,000$ data points for an application takes about 8.3 hours with a tuning period of 10 seconds.



\section{Conclusions}
Although there is much progress in building system software for memory tiering, tuning system parameters for memory tiering is largely missing. Tuning system parameters for memory tiering can bring large performance benefits but is challenging due to the concerns on runtime overhead, tuning convergence, and input-awareness. In this paper, we look deeply into representative memory-tiering solutions and introduce a novel tuning framework. We hope that our work can make memory tiering more easily deployable. 


\clearpage
\bibliographystyle{plain}
\bibliography{bib/sherry, bib/li, bib/su}

@inproceedings{nas,
    author = {Bailey, D. H. and Dagum, L. and Barszcz, E. and Simon, H.  D.},
    title = {NAS parallel benchmark results},
    booktitle = {Supercomputing '92: Proceedings of the 1992 ACM/IEEE
	conference on Supercomputing},
    year = {1992},
    isbn = {0-8186-2630-5},
    pages = {386--393},
    location = {Minneapolis, Minnesota, United States},
    publisher = {IEEE Computer Society Press},
    address = {Los Alamitos, CA, USA},
}

@misc{autonuma,
    author = "J. Corbet",
    title = "{{AutoNUMA:  the  Other  Approach  to  NUMA Scheduling}}",
    howpublished = "http://lwn.net/Articles/488709"
}

@misc{scikit-learn,
    author = "scikit-learn documentation",
    title = {{Tuning the Hyper-parameters of An Estimator”}},
    howpublished = "https://scikit-learn.org/stable/modules/grid_search.html"
}

@inproceedings{eurosys24:mtm,
    author = "Jie Ren and Dong Xu and Junhee Ryu and Kwangsik Shin and Daewoo Kim and Dong Li",
    title = {{MTM: Rethinking Memory Profiling and Migration for Multi-Tiered Large Memory Systems}}, 
    booktitle = "European Conference on Computer Systems",
    year = "2024"
}

@inproceedings{atc24_hm,
author = {Dong Xu and Junhee Ryu and Jinho Baek and Kwangsik Shin and Pengfei Su and Dong Li},
title = {{FlexMem: Adaptive Page Profiling and Migration for Tiered Memory}},
year = {2024},
booktitle = {30th USENIX Annual Technical Conference (ATC)}
}

@misc{amd_ibs,
    author = "{{AMD}}",
    title = "{{Analysis with Instruction Based Sampling}}",
    howpublished = "https://docs.amd.com/r/en-US/57368-uProf-user-guide",
year = {2025}
}

@misc{upm,
    author = "{{Anonymous}}",
    title = "{{User Space Page Migration}}",
    howpublished = "https://github.com/PTMT2024/UPM"
}

@misc{ptmt,
 author ={Anonymous},
 title={{PTMT}},
year = {2025},
 note={\url{https://anonymous.4open.science/r/PTMT}}
}

@article{graph500,
  title={{Introducing the graph 500}},
  author={Murphy, Richard C and Wheeler, Kyle B and Barrett, Brian W and Ang, James A},
  journal={Cray Users Group (CUG)},
  volume={19},
  number={45-74},
  pages={22},
  year={2010}
}

@misc{autonuma_tierd_memory,
    author = {Ying Huang},
    title = {autonuma: Optimize page placement for memory tiering system},
    year = {2020},
    note = {\url{https://patchwork.kernel.org/project/linux-mm/patch/20201027063217.211096-2-ying.huang@intel.com/}}
}

@misc{kmeans,
    title = {k-means clustering},
    note = {\url{https://en.wikipedia.org/wiki/K-means_clustering}},
    year =         2025
}

@misc{GUPS,
    title = {{GUPS (Giga Updates Per Second)}},
    note ={\url{https://icl.utk.edu/projectsfiles/hpcc/RandomAccess/}},
    year =         2025
}

@misc{cluster_elbow,
    title = {{Elbow Method}},
    note = {\url{https://www.scikit-yb.org/en/latest/api/cluster/elbow.html}},
    year =         2022
}

@inproceedings{tiering0.8,
  author =       {Vishal Verma.},
  title =        {{Tiering-0.8}},
  year =         2022,
  note =         {\url{https://git.kernel.org/pub/scm/linux/
kernel/git/vishal/tiering.git/log/?h=tiering-0.8}}
}

@inproceedings{meta_tpp,
  title={{TPP: Transparent page placement for CXL-enabled tiered-memory}},
  author={Maruf, Hasan Al and Wang, Hao and Dhanotia, Abhishek and Weiner, Johannes and Agarwal, Niket and Bhattacharya, Pallab and Petersen, Chris and Chowdhury, Mosharaf and Kanaujia, Shobhit and Chauhan, Prakash},
  booktitle={Proceedings of the 28th ACM International Conference on Architectural Support for Programming Languages and Operating Systems (ASPLOS)},
  year={2023}
}

@inproceedings{intel_pcm,
  title={{Intel® Performance Counter Monitor (Intel® PCM)}},
  author={Intel},
  note = {\url{https://github.com/intel/pcm}}
}

@inproceedings{HotBox2022bergman,
  title={{Reconsidering OS Memory Optimizations in the Presence of Disaggregated Memory}},
  author={Bergman, Shai and Faldu, Priyank and Grot, Boris and Vilanova, Llu{\'\i}s and Silberstein, Mark},
  booktitle={Proceedings of the 2022 ACM SIGPLAN International Symposium on Memory Management},
  year={2022}
}

@inproceedings{AutoTiering2021,
  title={Exploring the design space of page management for $\{$Multi-Tiered$\}$ memory systems},
  author={Kim, Jonghyeon and Choe, Wonkyo and Ahn, Jeongseob},
  booktitle={2021 USENIX Annual Technical Conference (USENIX ATC 21)},
  year={2021}
}

@inproceedings {Nomad2024,
author = {Lingfeng Xiang and Zhen Lin and Weishu Deng and Hui Lu and Jia Rao and Yifan Yuan and Ren Wang},
title = {{Nomad: Non-Exclusive Memory Tiering via Transactional Page Migration}},
booktitle = {USENIX Symposium on Operating Systems Design and Implementation (OSDI)},
year = {2024}
}

@INPROCEEDINGS{TMP2021,
  author={Choi, Jinyoung and Blagodurov, Sergey and Tseng, Hung-Wei},
  booktitle={2021 IEEE International Parallel and Distributed Processing Symposium (IPDPS)}, 
  title={Dancing in the Dark: Profiling for Tiered Memory}, 
  year={2021}
}

@inproceedings{Hemem2021,
  title={Hemem: Scalable tiered memory management for big data applications and real nvm},
  author={Raybuck, Amanda and Stamler, Tim and Zhang, Wei and Erez, Mattan and Peter, Simon},
  booktitle={Proceedings of the ACM SIGOPS 28th Symposium on Operating Systems Principles},
  year={2021}
}

@inproceedings{MaPHeA2021,
  title={MaPHeA: A lightweight memory hierarchy-aware profile-guided heap allocation framework},
  author={Oh, Deok-Jae and Moon, Yaebin and Lee, Eojin and Ham, Tae Jun and Park, Yongjun and Lee, Jae W and Ahn, Jung Ho},
  booktitle={Proceedings of the 22nd ACM SIGPLAN/SIGBED International Conference on Languages, Compilers, and Tools for Embedded Systems},
  year={2021}
}

@inproceedings{Memtis2023,
  title={Memtis: Efficient memory tiering with dynamic page classification and page size determination},
  author={Lee, Taehyung and Monga, Sumit Kumar and Min, Changwoo and Eom, Young Ik},
  booktitle={Proceedings of the 29th Symposium on Operating Systems Principles},
  year={2023}
}

@article{heterovisor2015,
  title={Heterovisor: Exploiting resource heterogeneity to enhance the elasticity of cloud platforms},
  author={Gupta, Vishal and Lee, Min and Schwan, Karsten},
  journal={ACM SIGPLAN Notices},
  volume={50},
  number={7},
  pages={79--92},
  year={2015},
  publisher={ACM New York, NY, USA}
}

@inproceedings{heteroos2017,
  title={Heteroos: Os design for heterogeneous memory management in datacenter},
  author={Kannan, Sudarsun and Gavrilovska, Ada and Gupta, Vishal and Schwan, Karsten},
  booktitle={Proceedings of the 44th Annual International Symposium on Computer Architecture},
  year={2017}
}

@INPROCEEDINGS{MULTI-CLOCK2022,
  author={Maruf, Adnan and Ghosh, Ashikee and Bhimani, Janki and Campello, Daniel and Rudoff, Andy and Rangaswami, Raju},
  booktitle={2022 IEEE International Symposium on High-Performance Computer Architecture (HPCA)}, 
  title={MULTI-CLOCK: Dynamic Tiering for Hybrid Memory Systems}, 
  year={2022}
}

@inproceedings{daos2022,
  title={Daos: Data access-aware operating system},
  author={Park, SeongJae and Bhowmik, Madhuparna and Uta, Alexandru},
  booktitle={Proceedings of the 31st International Symposium on High-Performance Parallel and Distributed Computing},
  year={2022}
}

@inproceedings{DAMON,
  title={{LRU-list manipulation with DAMON}},
  author={Jonathan Corbet},
  note = {\url{https://lwn.net/Articles/905370/}},
  year = {2022}
}

@inproceedings{doudali2019kleio,
  title={Kleio: A hybrid memory page scheduler with machine intelligence},
  author={Doudali, Thaleia Dimitra and Blagodurov, Sergey and Vishnu, Abhinav and Gurumurthi, Sudhanva and Gavrilovska, Ada},
  booktitle={Proceedings of the 28th International Symposium on High-Performance Parallel and Distributed Computing},
  year={2019}
}

@inproceedings{doudali2021cori,
  title={Cori: Dancing to the right beat of periodic data movements over hybrid memory systems},
  author={Doudali, Thaleia Dimitra and Zahka, Daniel and Gavrilovska, Ada},
  booktitle={2021 IEEE International Parallel and Distributed Processing Symposium (IPDPS)},
  year={2021}
}

@inproceedings{doudali2022coeus,
  title={Coeus: Clustering (a) like patterns for practical machine intelligent hybrid memory management},
  author={Doudali, Thaleia Dimitra and Gavrilovska, Ada},
  booktitle={2022 22nd IEEE International Symposium on Cluster, Cloud and Internet Computing (CCGrid)},
  year={2022}
}

@inproceedings{doudali2022cronus,
  title={Cronus: Computer Vision-based Machine Intelligent Hybrid Memory Management},
  author={Doudali, Thaleia Dimitra and Gavrilovska, Ada},
  booktitle={Proceedings of the 2022 International Symposium on Memory Systems},
  year={2022}
}

@inproceedings{wu2017unimem,
  title={Unimem: Runtime data managementon non-volatile memory-based heterogeneous main memory},
  author={Wu, Kai and Huang, Yingchao and Li, Dong},
  booktitle={Proceedings of the International Conference for High Performance Computing, Networking, Storage and Analysis},
  year={2017}
}

@inproceedings{vasilakis2020hybrid2,
  title={Hybrid2: Combining caching and migration in hybrid memory systems},
  author={Vasilakis, Evangelos and Papaefstathiou, Vassilis and Trancoso, Pedro and Sourdis, Ioannis},
  booktitle={2020 IEEE International Symposium on High Performance Computer Architecture (HPCA)},
  year={2020}
}

@inproceedings{baliosian2009rule,
  title={A rule-based distributed system for self-optimization of constrained devices},
  author={Baliosian, Javier and Visca, Jorge and Grampin, Eduardo and Vidal, Leonardo and Giachino, Martin},
  booktitle={2009 IFIP/IEEE International Symposium on Integrated Network Management},
  year={2009}
}

@inproceedings{krish2014hats,
  title={hats: A heterogeneity-aware tiered storage for hadoop},
  author={Krish, KR and Anwar, Ali and Butt, Ali R},
  booktitle={2014 14th IEEE/ACM International Symposium on Cluster, Cloud and Grid Computing},
  year={2014}
}

@inproceedings{servat2017automating,
  title={Automating the application data placement in hybrid memory systems},
  author={Servat, Harald and Pe{\~n}a, Antonio J and Llort, Germ{\'a}n and Mercadal, Estanislao and Hoppe, Hans-Christian and Labarta, Jes{\'u}s},
  booktitle={2017 IEEE International Conference on Cluster Computing (CLUSTER)},
  year={2017}
}

@inproceedings{dulloor2016data_tiering,
  title={Data tiering in heterogeneous memory systems},
  author={Dulloor, Subramanya R and Roy, Amitabha and Zhao, Zheguang and Sundaram, Narayanan and Satish, Nadathur and Sankaran, Rajesh and Jackson, Jeff and Schwan, Karsten},
  booktitle={Proceedings of the Eleventh European Conference on Computer Systems},
  year={2016}
}

@inproceedings{ni2023tmc,
  title={TMC: Near-Optimal Resource Allocation for Tiered-Memory Systems},
  author={Ni, Yuanjiang and Mehra, Pankaj and Miller, Ethan and Litz, Heiner},
  booktitle={Proceedings of the 2023 ACM Symposium on Cloud Computing},
  year={2023}
}

@inproceedings{Google_TMTS2023,
  title={Towards an adaptable systems architecture for memory tiering at warehouse-scale},
  author={Duraisamy, Padmapriya and Xu, Wei and Hare, Scott and Rajwar, Ravi and Culler, David and Xu, Zhiyi and Fan, Jianing and Kennelly, Christopher and McCloskey, Bill and Mijailovic, Danijela and others},
  booktitle={Proceedings of the 28th ACM International Conference on Architectural Support for Programming Languages and Operating Systems, Volume 3},
  year={2023}
}

@inproceedings{Google_farmemory2019,
  title={Software-defined far memory in warehouse-scale computers},
  author={Lagar-Cavilla, Andres and Ahn, Junwhan and Souhlal, Suleiman and Agarwal, Neha and Burny, Radoslaw and Butt, Shakeel and Chang, Jichuan and Chaugule, Ashwin and Deng, Nan and Shahid, Junaid and others},
  booktitle={Proceedings of the Twenty-Fourth International Conference on Architectural Support for Programming Languages and Operating Systems},
  year={2019}
}

@inproceedings{Google_Llama2020,
  title={Learning-based memory allocation for C++ server workloads},
  author={Maas, Martin and Andersen, David G and Isard, Michael and Javanmard, Mohammad Mahdi and McKinley, Kathryn S and Raffel, Colin},
  booktitle={Proceedings of the Twenty-Fifth International Conference on Architectural Support for Programming Languages and Operating Systems},
  year={2020}
}

@inproceedings{idt2024chang,
  title={{IDT: Intelligent Data Placement for Multi-tiered Main Memory with Reinforcement Learning}},
  author={Chang, Juneseo and Doh, Wanju and Moon, Yaebin and Lee, Eojin and Ahn, Jung Ho},
  booktitle={International Symposium on High-Performance Parallel and Distributed Computing (HPDC)},
  year={2024}
}

@article{AMP2020,
  title={Adaptive page migration policy with huge pages in tiered memory systems},
  author={Heo, Taekyung and Wang, Yang and Cui, Wei and Huh, Jaehyuk and Zhang, Lintao},
  journal={IEEE Transactions on Computers},
  volume={71},
  number={1},
  pages={53--68},
  year={2020},
  publisher={IEEE}
}

@article{MMDTune2023Ye,
  title={Parameters tuning of multi-model database based on deep reinforcement learning},
  author={Ye, Feng and Li, Yang and Wang, Xiwen and Nedjah, Nadia and Zhang, Peng and Shi, Hong},
  journal={Journal of Intelligent Information Systems},
  volume={61},
  number={1},
  pages={167--190},
  year={2023},
  publisher={Springer}
}

@article{MMDTune+2023Sun,
  title={Workload-Aware Performance Tuning for Multimodel Databases Based on Deep Reinforcement Learning},
  author={Sun, Jun and Ye, Feng and Nedjah, Nadia and Zhang, Ming and Xu, Dong},
  journal={International Journal of Intelligent Systems},
  volume={2023},
  number={1},
  pages={8835111},
  year={2023},
  publisher={Wiley Online Library}
}

@inproceedings{CDBTune2019,
  title={An end-to-end automatic cloud database tuning system using deep reinforcement learning},
  author={Zhang, Ji and Liu, Yu and Zhou, Ke and Li, Guoliang and Xiao, Zhili and Cheng, Bin and Xing, Jiashu and Wang, Yangtao and Cheng, Tianheng and Liu, Li and others},
  booktitle={Proceedings of the 2019 international conference on management of data},
  year={2019}
}

@inproceedings{OtterTune2017,
  title={Automatic database management system tuning through large-scale machine learning},
  author={Van Aken, Dana and Pavlo, Andrew and Gordon, Geoffrey J and Zhang, Bohan},
  booktitle={Proceedings of the 2017 ACM international conference on management of data},
  year={2017}
}

@article{MPOD2023,
  title={An energy-efficient tuning method for cloud servers combining DVFS and parameter optimization},
  author={Lin, Weiwei and Luo, Xiaoxuan and Li, ChunKi and Liang, Jiechao and Wu, Guokai and Li, Keqin},
  journal={IEEE Transactions on Cloud Computing},
  year={2023},
  publisher={IEEE}
}

@inproceedings{einziger2018cache,
  title={Adaptive software cache management},
  author={Einziger, Gil and Eytan, Ohad and Friedman, Roy and Manes, Ben},
  booktitle={Proceedings of the 19th International Middleware Conference},
  year={2018}
}

@article{Qtune2019,
  title={Qtune: A query-aware database tuning system with deep reinforcement learning},
  author={Li, Guoliang and Zhou, Xuanhe and Li, Shifu and Gao, Bo},
  journal={Proceedings of the VLDB Endowment},
  volume={12},
  number={12},
  pages={2118--2130},
  year={2019},
  publisher={VLDB Endowment}
}

@inproceedings{CAPES2017,
  title={CAPES: Unsupervised storage performance tuning using neural network-based deep reinforcement learning},
  author={Li, Yan and Chang, Kenneth and Bel, Oceane and Miller, Ethan L and Long, Darrell DE},
  booktitle={Proceedings of the international conference for high performance computing, networking, storage and analysis},
  year={2017}
}

@inproceedings{autosys2020,
  title={$\{$AutoSys$\}$: The Design and Operation of $\{$Learning-Augmented$\}$ Systems},
  author={Liang, Chieh-Jan Mike and Xue, Hui and Yang, Mao and Zhou, Lidong and Zhu, Lifei and Li, Zhao Lucis and Wang, Zibo and Chen, Qi and Zhang, Quanlu and Liu, Chuanjie and others},
  booktitle={2020 USENIX Annual Technical Conference (USENIX ATC 20)},
  year={2020}
}

@inproceedings{Config-Snob2024,
  title={$\{$Config-Snob$\}$: Tuning for the Best Configurations of Networking Protocol Stack},
  author={Bin-Yahya, Manaf and Zhao, Yifei and Shafieirad, Hossein and Ho, Anthony and Yin, Shijun and Wang, Fanzhao and Li, Geng},
  booktitle={2024 USENIX Annual Technical Conference (USENIX ATC 24)},
  year={2024}
}

@inproceedings{ren2019archivist,
  title={Archivist: A machine learning assisted data placement mechanism for hybrid storage systems},
  author={Ren, Jinting and Chen, Xianzhang and Tan, Yujuan and Liu, Duo and Duan, Moming and Liang, Liang and Qiao, Lei},
  booktitle={2019 IEEE 37th International Conference on Computer Design (ICCD)},
  year={2019}
}

@inproceedings{cheng2019optimizing,
  title={Optimizing data placement on hierarchical storage architecture via machine learning},
  author={Cheng, Peng and Lu, Yutong and Du, Yunfei and Chen, Zhiguang and Liu, Yang},
  booktitle={Network and Parallel Computing: 16th IFIP WG 10.3 International Conference},
  year={2019}
}

@inproceedings{colloid2024,
  title={Tiered Memory Management: Access Latency is the Key!},
  author={Vuppalapati, Midhul and Agarwal, Rachit},
  booktitle={Proceedings of the ACM SIGOPS 30th Symposium on Operating Systems Principles},
  year={2024}
}

@inproceedings{song2025hybridtier,
  title={HybridTier: an Adaptive and Lightweight CXL-Memory Tiering System},
  author={Song, Kevin and Yang, Jiacheng and Wang, Zixuan and Zhao, Jishen and Liu, Sihang and Pekhimenko, Gennady},
  booktitle={Proceedings of the 30th ACM International Conference on Architectural Support for Programming Languages and Operating Systems (ASPLOS)},
  year={2025}
}

@inproceedings{huaicheng2025tiered,
  title={Tiered Memory Management Beyond Hotness},
  author={Liu, Jinshu and Hadian, Hamid and Xu, Hanchen and Li, Huaicheng},
  booktitle={19th USENIX Symposium on Operating Systems Design and Implementation (OSDI 25)},
  year={2025}
}

@misc{NASA_NPB,
    title = {{NAS Parallel Benchmarks}},
    note ={\url{https://www.nas.nasa.gov/software/npb.html}},
    year =         2024
}

@misc{stable-baselines3,
    title = {{Stable Baselines3}},
    note ={\url{https://github.com/DLR-RM/stable-baselines3}},
    year =         2025
}

@misc{normal_dist,
    title = {{Normal distribution}},
    note ={\url{https://en.wikipedia.org/wiki/Normal_distribution}},
    year =         2025
}

@INPROCEEDINGS{feature_weighted_knn,
  author={Mladenova, Tsvetelina},
  booktitle={International Symposium on Multidisciplinary Studies and Innovative Technologies (ISMSIT)}, 
  title={A Feature-Weighted Rule for the K-Nearest Neighbor}, 
  year={2021}
}

@article{RL_PPO,
  title={Proximal policy optimization algorithms},
  author={Schulman, John and Wolski, Filip and Dhariwal, Prafulla and Radford, Alec and Klimov, Oleg},
  journal={arXiv preprint arXiv:1707.06347},
  year={2017}
}

@article{behavioral_cloning,
  title={Behavioral cloning from observation},
  author={Torabi, Faraz and Warnell, Garrett and Stone, Peter},
  journal={arXiv preprint arXiv:1805.01954},
  year={2018}
}

@inproceedings{silo,
  title={Speedy transactions in multicore in-memory databases},
  author={Tu, Stephen and Zheng, Wenting and Kohler, Eddie and Liskov, Barbara and Madden, Samuel},
  booktitle={Proceedings of the Twenty-Fourth ACM Symposium on Operating Systems Principles},
  year={2013}
}

@article{liblinear,
  title={LIBLINEAR: A library for large linear classification},
  author={Fan, Rong-En and Chang, Kai-Wei and Hsieh, Cho-Jui and Wang, Xiang-Rui and Lin, Chih-Jen},
  journal={Journal of machine Learning research},
  volume={9},
  pages={1871--1874},
  year={2008}
}

@misc{PTMT_UPM,
    title = {{User-Space Page Management}},
    note ={\url{https://anonymous.4open.science/r/UPM}},
    year =         2025
}

@inproceedings{wu2015enabling,
  title={Enabling and exploiting flexible task assignment on GPU through SM-centric program transformations},
  author={Wu, Bo and Chen, Guoyang and Li, Dong and Shen, Xipeng and Vetter, Jeffrey},
  booktitle={Proceedings of the 29th ACM on International Conference on Supercomputing},
  year={2015}
}

@article{eyerman2008system,
  title={System-level performance metrics for multiprogram workloads},
  author={Eyerman, Stijn and Eeckhout, Lieven},
  journal={IEEE micro},
  volume={28},
  number={3},
  pages={42--53},
  year={2008},
  publisher={IEEE}
}

@misc{aws-hpc-cfd,
    title = {{Computational Fluid Dynamics}},
    note  = {\url{https://aws.amazon.com/hpc/cfd}},
    year  = {2025}
}

@misc{aws-hpc-hcls,
    title = {{High Performance Computing for Healthcare \& Life Sciences}},
    note  = {\url{https://aws.amazon.com/hpc/hcls}},
    year  = {2025}
}

@misc{gcp-cae,
    title = {{Running Computer-Aided Engineering Workloads on Google Cloud}},
    note  = {\url{https://cloud.google.com/solutions/running-computer-aided-engineering-workloads}},
    year  = {2024}
}

@misc{rescale-hpc-saas,
    title = {{HPC as a Service}},
    note  = {\url{https://rescale.com/platform/hpc-as-a-service/}},
    year  = {2025}
}

@misc{simscale-saas,
    title = {{AI-Native Engineering Simulation in the Cloud}},
    note  = {\url{https://www.simscale.com/}},
    year  = {2025}
}

@misc{azure-model-catalog,
    title = {{Azure AI Foundry Model Catalog}},
    note  = {\url{https://azure.microsoft.com/en-us/products/ai-foundry/models}},
    year  = {2025}
}

@misc{vertex-model-garden,
    title = {{Model Garden on Vertex AI}},
    note  = {\url{https://cloud.google.com/model-garden}},
    year  = {2024}
}

@inproceedings{TMO_meta,
  title={TMO: Transparent memory offloading in datacenters},
  author={Weiner, Johannes and Agarwal, Niket and Schatzberg, Dan and Yang, Leon and Wang, Hao and Sanouillet, Blaise and Sharma, Bikash and Heo, Tejun and Jain, Mayank and Tang, Chunqiang and others},
  booktitle={Proceedings of the 27th ACM International Conference on Architectural Support for Programming Languages and Operating Systems},
  year={2022}
}

@inproceedings{cmpi2025,
  title={cMPI: Using CXL Memory Sharing for MPI One-Sided and Two-Sided Inter-Node Communications},
  author={Wang, Xi and Ma, Bin and Kim, Jongryool and Koh, Byungil and Kim, Hoshik and Li, Dong},
  booktitle={Proceedings of the International Conference for High Performance Computing, Networking, Storage and Analysis},
  year={2025}
}

@article{xu2026cccl,
  title={CCCL: Node-Spanning GPU Collectives with CXL Memory Pooling},
  author={Xu, Dong and Meng, Han and Chen, Xinyu and Zhu, Dengcheng and Tang, Wei and Liu, Fei and Xie, Liguang and Xiang, Wu and Shi, Rui and Li, Yue and others},
  journal={arXiv preprint arXiv:2602.22457},
  year={2026}
}

@inproceedings{wang2025performance,
  title={Performance Characterization of CXL Memory and Its Use Cases},
  author={Wang, Xi and Liu, Jie and Wu, Jianbo and Yang, Shuangyan and Ren, Jie and Shankar, Bhanu and Li, Dong},
  booktitle={2025 IEEE International Parallel and Distributed Processing Symposium (IPDPS)},
  pages={1048--1061},
  year={2025},
  organization={IEEE}
}

@inproceedings{ren2025machine,
  title={Machine learning-guided memory optimization for DLRM inference on tiered memory},
  author={Ren, Jie and Ma, Bin and Yang, Shuangyan and Francis, Benjamin and Ardestani, Ehsan K and Si, Min and Li, Dong},
  booktitle={2025 IEEE International Symposium on High Performance Computer Architecture (HPCA)},
  pages={1631--1647},
  year={2025},
  organization={IEEE}
}
\end{document}